\documentstyle[preprint,aps,prb,amssymb]{revtex}
\newcommand{\be}{\begin{equation}}
\newcommand{\ee}{\end{equation}}
\newcommand{\bq}{\begin{eqnarray}}
\newcommand{\eq}{\end{eqnarray}}

\def\elle{{\it l}}
\def\dsp{\displaystyle}
\def\nis{n_{{\bf i}, \sigma}}
\def\cisd{c_{{\bf i}, \sigma}^{\dagger}}
\def\cis{c_{{\bf i}, \sigma}}
\def\cjs{c_{{\bf j}, \sigma}}

\def\nks{n_{{\bf k},\sigma}}
\def\nup{n_{{\bf i}, \uparrow}}
\def\ndw{n_{{\bf i}, \downarrow}}
\def\si{\sum_{\bf i}}
\def\sij{\sum_{<{\bf i},{\bf j}>}}

\def\diracij{\delta_{{\bf i}, {\bf j}}}
\def\sigsig{\delta_{\sigma , \, \sigma '}}

\begin{document}
\draft
\title{Dynamics of the Hubbard Model: a general approach
 by Time-Dependent Variational Principle}   
\author{Arianna Montorsi, and Vittorio Penna} 
\address{
Dipartimento di Fisica and Unit\'a INFM, Politecnico di Torino,  
I-10129 Torino, Italy}
\date{30 April 1996}
\maketitle

\begin{abstract}
We describe the quantum dynamics of the Hubbard model at semi-classical
level, by implementing the Time-Dependent Variational Principle (TDVP)
procedure on appropriate macroscopic wavefunctions constructed in terms of
$su(2)$-coherent states. Within the TDVP procedure, such states turn out to
include a time-dependent quantum phase, part of which can be recognized as
Berry's phase. We derive two new semi-classical model Hamiltonians for
describing the dynamics in the paramagnetic, superconducting, 
antiferromagnetic and charge density wave phases and solve the corresponding 
canonical equations of motion in various cases. Noticeably, a vortex-like 
ground state phase dynamics is found to take place for $U>0$ away from 
half filling. Moreover, it appears that an oscillatory-like ground state 
dynamics survives at the Fermi surface at half-filling for any $U$. The 
low-energy dynamics is also exactly solved by separating fast and slow 
variables. The role of the time-dependent phase is shown to be particularly 
interesting in the ordered phases. 
\end{abstract}

\pacs{71.10.Fd,03.65.s,03.65.B,47.32.Cc}

\narrowtext
\section{Introduction}
Interest in strongly-correlated itinerant electron systems has been
constantly growing in the last three decades. Especially since the
discovery (almost ten years ago) of high-$T_c$ superconductors,
an enormous amount of work has been devoted to such systems, aimed both
to investigate their macroscopic thermodynamical properties via experimental
mesurments, and to disclose --by employing the standard methods of statistical
mechanics-- what type of macroscopic collective order is responsable for the
frictionless regime.

Nevertheless, due to the high number of variables naturally involved in the
models proposed for investigating these many-electron systems, and probably to
the background of the community of physicists who first considered these models,
to the best of our knowledge very little effort has been made in order to
investigate their dynamical behavior. On the other hand, this type of analysis
is known to lead to interesting properties of superfluidity when applied
for instance to the BCS Hamiltonian \cite{BMT}. 

Two circumstances, at least, prompt to attempt the dynamical approach and to
carefully consider its possible developments. First of all,
standard theoretical techniques such as the Time-Dependent Variational
Principle (TDVP) procedure and its path-integral version, the Stationary Phase
Approximation method, have been remarkably developed in the recent years,  
by exploiting the notion of Generalized Coherent State (GCS)\cite{PER} and the
spectrum generating (or dynamical) algebra method. At the formal level, such
group-theoretical
tools have greatly simplified and provided of a systematic character the
TDVP procedure, which essentially consists in reducing the system quantum
dynamics to a
semiclassical hamiltonian form. The procedure, formulated, for example, in
the form of refs. \cite{ZFG} and \cite{FUTS}, was first introduced for
studying the low-lying collective states in nuclei, but it is easily extended
also to any systems endowed with a large number of degrees of freedom.

Furthermore the special role assigned to the quantum phase of the macroscopic
trial wavefunction involved within the TDVP framework makes the procedure even
more attractive. Such quantum phase, in fact, is structured so as to have a
memory of the entire dynamical behavior. It is thus natural expecting some
kind of correlation between the type of microscopic order which possibly
characterizes the medium and the phase time-behavior. Such kind of
effects have been for instance investigated in \cite{BRLO}, where the
study of quantum dynamics of solitons in ferromagnets was shown to have
remarkable consequences on Berry phase behavior. 

This aspect, in turn, directly leads to the second circumstance which
motivates our interest for the dynamical viewpoint. At low temperature,
one can reasonably think of performing current measurements or
superconductive quantum interference measurments based on an experimental
devices similar to the ones employed to investigate the $\it Josephson$
$\it effect$ \cite{TITI,HOHA}. It is essential recalling that, in that case,
the time-dependence of the order
paramenter (the macroscopic wavefunction of the system) has a crucial role in
allowing for the detection of microscopic phenomena occuring in the medium.
Although the TDVP approach is able of taking into account large number
of dynamical degrees of freedom, and is thus fitting to describe a
strongly-correlated electron gas, one can expect that low excited states
actually involve a restricted number of dynamical variables. Under such
conditions the quantum phase could retain a nonrandom character which
makes it suitable for experimental measurments.

A further quality of the TDVP approach must be still pointed out. Such scheme,
in fact, involves the construction of new semiclassical Hamiltonians which
are obtained in a completely independent way with respect to the model
Hamiltonians derived by standard mean-field techniques of statistical mechanics
(for instance Hartree-Fock). The ground state of the semiclassical Hamiltonian
reproduces, as we shall see, the same results of the
Hartree-Fock approach from the set of dynamical fixed points. Moreover, 
as opposite to the mean-field cases, here also the excited states at low
energy are expected to be a realistic description of those of the
original model Hamiltonian. This character is related with the fact that
TDVP Hamiltonians, even though affected by the approximations imposed by
the method, generally preserve a structure rather faithful to the
second quantized Hamiltonian. Comparison with statistical mechanics
approximate models is thus interesting in any case.

In this paper we propose the implementation of an appropriate
generalization of TDVP to itinerant interacting electron systems.
This amounts to applying the TDVP
to a trial wave-function representing a semiclassical macroscopic state
constructed by generalized coherent states of the dynamical algebra of the
model Hamiltonian \cite{PER,ZFG}. From the semiclassical picture of the
system obtained in this way one can derive canonical equations of
motion, and a classical description of system's dynamics.
A key role within this approach is played by a time-dependent phase,
which has to be fixed so as to satisfy --at least in average-- the
Schr\"odinger equation. In \cite{FUTS} it was  shown that, under appropriate
assumptions, the latter is nothing but the dynamical plus geometric
phases\cite{SHWI} beyond the adiabatic approximation.

Here we apply the above method to the Hubbard model\cite{HUB}, described
by the Hamiltonian
\bq
H_{Hub} &=& - T\sij\sum_\sigma \cisd \cjs +  U \si \nup\ndw
-\mu\si(\nup+\ndw)\nonumber \\
&=& \sum_{{\bf k},\sigma} (\epsilon_{\bf k}-\mu) \nks + U \sum_{{\bf k},
{\bf l}, {\bf m}} a_{{\bf k}, \uparrow}^\dagger a_{{\bf m},
\downarrow}^\dagger a_{{\bf l}, \downarrow} a_{{\bf k}+{\bf m}-
{\bf l}, \uparrow} \quad ,\label{hamhub}  
\eq
where, on the first line, $\cisd , \cjs \,$ are fermionic creation
and annihilation operators ($
\{ \cjs, c_{{\bf i},\sigma'} \} = 0 \,$, $\, \{ \cjs ,
c_{{\bf i},\sigma'}^\dagger \} = \diracij\, \sigsig
{{\Bbb I}}$, $\, \nis \doteq \cisd \cis$) on a $d$-dimensional
lattice $\Lambda$ (${\bf i} , \, {\bf j} \in \Lambda$, $\sigma \in \{
\Uparrow , \Downarrow \}$) with $N$ sites, and $< {\bf i} , \, {\bf j} >$
stands for
nearest neighbors ({\it n.n.}) in $\Lambda$. In the second line the
same Hamiltonian is rewritten in the reciprocal space $\tilde\Lambda$,
with $\dsp{a_{{\bf k}, \sigma}\doteq \sum_{\bf j}
{\rm e}^{i \pi {\bf j}\cdot{\bf k}} \cjs}$, $\dsp{\epsilon_{\bf k}\doteq
- 2 T\sum_{r=1}^d \cos{k_r}}$.
In (\ref{hamhub}) the first term represents the tight-binding band energy of the
electrons ($T$ being the hopping amplitude), the $U$ term describes
their on-site Coulomb interaction, and $\mu$ is the chemical potential,
which will allow us to fix the conserved quantity $ {\cal N}_e =
\si(\nup+\ndw)$, {\sl i.e.} the total electron number operator on the lattice.

Being the GCS of the dynamical algebra of Hamiltonian (\ref{hamhub}) quite
complex to deal
with, we choose as trial GCS's for constructing the semiclassical
macroscopic state the $su(2)$ coherent states which are exact for the
corresponding Hartree-Fock Hamiltonian.
This is done for two different cases, namely that 
describing superconducting (SC) and paramagnetic phases, and, at
half-filling, that describing antiferromagnetic (AF) and
charge-density-wave (CDW) phases.
The approximate equations of motion we obtain for the
full Hamiltonian in the two cases are then solved in
some integrable cases, and by approximate methods in other
interesting limits. This gives rise to a variety of different dynamical
behaviors, from vortex-like dynamics in the ground state 
for the $U>0$, $n\neq1$ regime, to oscillations of the number of electrons
around the Fermi surface and possible laser effect at low energy, and to
single-mode collective frequency dynamics, which should reflect the occurence of
macroscopic order in the medium. In particular, 
the time-dipendent phase, which -- due to its macroscopic nature -- can
be considered as an observable quantity, exhibit a behavior which is
shown to be related to the non-vanishing of order parameters, and is evaluated
exactly in several situations.

The paper is organized as follows. In section II we review the
generalized TDVP approach and its connections with the quantum
geometric phase. In section III we treat explicitly the SC-paramagnetic case,
by constructing first the macroscopic trial wave-function and the
corresponding semiclassical Hamiltonian, and deriving then the canonical
equations of motion with the time-dependent phase factor.
Section IV is devoted to study the fixed points of these equations and in
particular the ground state solutions and metastable states; obtaining 
for the ground state the Hartree-Fock results  as well as 
a nontrivial vortex-like dynamics in the so called paramagnetic phase,
and topological excitations for the metastable states.
In section V we analyze an integrable case, which exhibits collective
order and non-zero pairing induced by the $k$-modes interactions.
In section VI we investigate the global dynamics by comparing slow with fast
degrees of freedom. We show how slow variables tend to constitute 
an autonomous subsystem which drives fast variables dynamics on
large time-scales. An integrable case where the slow subsystem is reduced to
a two-level system representing $k$-modes close to the Fermi level, is
explicitly solved in section VII.
In section VIII we repeat some of the above analysis for the AF phase 
at half-filling, finding in particular an oscillating behavior  
at the Fermi surface. The final section is devoted to some
conclusions.

\section{Generalized TDVP method}

Knowledge of the dynamical algebra ${\cal G}$ of a given (time-independent)
Hamiltonian $H$ allows the construction of an over-complete set of states
known as generalized coherent states,
\bq
\vert \Phi_0> ={\rm exp}\left (\sum_\alpha[\theta_\alpha E_\alpha -
\theta_\alpha^* E_{-\alpha}]\right )\vert 0> \nonumber \quad ,
\eq
where $\{E_\alpha, E_{-\alpha}\}$ are the raising and
lowering operators in the Cartan representation of ${\cal G}$,
and $\vert 0>$ is the highest weight of the representation, defined
by $E_{-\alpha}\vert 0>=0$ for all positive $\alpha$'s.
The state $\vert \Phi(t)>$, which is the obvious
time-dependent generalization of $\vert \Phi_0>$, 
\bq
\vert \Phi(t)>
= {\rm exp}\left (\sum_\alpha[\theta_\alpha(t) E_\alpha -
\theta_\alpha(t)^* E_{-\alpha}]\right )\vert 0> \quad , \nonumber
\eq
is related to the time evolution of $\vert \Phi_0>$, described by
the state $\vert \Psi(t)>\doteq\dsp{{\rm e}^{-{i\over\hbar} H t}\vert \Phi_0>}$,
through
\be
\vert \Psi(t)>\equiv{\rm e}^{i {\varphi (t)\over \hbar}}\vert \Phi(t)> \quad , 
\label{psi1}
\ee
where
\be
\varphi\doteq -{\cal H}t +i\hbar\int_0^t d\tau<\Phi(\tau)\vert 
{\partial\over{\partial \tau}}\vert \Phi(\tau)>\quad ,\label{tdp}
\ee
and ${\cal H}=<\Phi(t)\vert H\vert \Phi(t)>$.

The time-dependence of the parameters $\theta_\alpha(t)$'s
is determined by imposing that $\vert \Psi(t)>$, as given by (\ref{psi1}),
satisfies the Schr\"odinger equation. It turns out that this
amounts to requiring the $\theta_\alpha(t)$'s obey the
canonical equation of motion (see below). 
An alternative parametrization for the state $\vert \Phi(t)>$
can be used, namely
\be
\vert \Phi(t)> = {\cal N}^{-{1\over 2}} {\rm exp} \left(\sum_{\alpha >0}
z_\alpha(t) E_\alpha\right) \vert 0> 
\quad , \label{phi}
\ee
${\cal N}$ being the normalization factor, such that $<\Phi(t)\vert \Phi(t)>=1$.
In this case the canonical equations of motion read\cite{FUTS}
\be
i \hbar \sum_\beta g_{\alpha,\beta} \dot z_\beta =
\frac{\partial {\cal H}}{\partial z^*_\alpha} \quad ,\quad
i \hbar \sum_\beta g^*_{\alpha,\beta} \dot z^*_\beta =
-\frac{\partial {\cal H}}{\partial z_\alpha} \quad , \label{caneq}
\ee
where $g_{\alpha,\beta}=\dsp{\frac{\partial^2\log {\cal N}}{\partial
z^*_\alpha \partial z_\beta}}$ is the metric of the phase space spanned
by $\{z_\alpha, z^*_\alpha\}$, and determines its symplectic structure. 
The metric $g_{\alpha,\beta}$ indeed determines the explicit form of Poisson's
brackets,
\be
\{A,B\}_{P.B.}=\sum_{\alpha,\beta} i\hbar (g^{-1})_{\alpha,\beta}
\left (\frac{\partial A}{\partial z^*_\alpha}
\frac{\partial B}{\partial z_\beta}-\frac{\partial A}{\partial
z_\beta} \frac{\partial B}{\partial z^*_\alpha}\right ) \quad ,
\label{pb}
\ee
where $g^{-1}$ represents the inverse matrix of $g$.

If the algebra ${\cal G}$ is the full dynamical algebra of $H$ 
({\sl i.e.} if $H\in{\cal G}$ for any choice of the physical parameters)
the above procedure is exact. In particular, it gives the exact quantum
ground state of $H$ as fixed point of the equations (\ref{caneq}).  
Nevertheless, in a many-body problem like the one described
by the Hamiltonian (\ref{hamhub}) the dimension of the dynamical
algebra is exponentially growing with $N$, and infinite in the
thermodynamical limit. Even though one may still work out the canonical
equations of motion\cite{MOPE}, their explicit
solution becomes then quite hard to handle. It is therefore reasonable
to inquire to which extent the above scheme can be used in the case where
the GCS are built
in a sub-algebra ${\cal A}\subset {\cal G}$ for which the equations
of motion become tractable. In this case of course $\vert \Psi(t)>$, as
given by (\ref{psi1}), differs from ${\rm e}^{-{\bar i\over\hbar}H
t}\vert \Phi_0>$, and the Schr\"odinger equation is in general not satisfied.
Nevertheless, the answer given by the generalized TDVP
approach\cite{FUTS} is that in fact the above scheme still holds also
in this case, if one just requires that at least the inner
product of the Schr\"odinger equation for $\vert \Psi(t)>$ with $<\Psi(t)\vert $
vanishes, {\sl i.e.}
\be
<\Psi(t)\vert \left (i\hbar\frac{\partial}{\partial t}- H\right)
\vert \Psi(t)> = 0 \quad . \label{schr}
\ee
Notice that now $\vert \Psi(t)>$ is to be built only with the raising
operators $E_\alpha\in {\cal A}$. Hence the analogy with the
exact solution (\ref{psi1})--(\ref{caneq})
(where $E_\alpha\in {\cal G}$)  is complete at a
formal level, but approximate in the results, 
${\cal H}$ being evaluated on a sub-space of the whole dynamical
algebra ${\cal G}$. Nevertheless, as for the reliability of the
method, one should recall that in the limit
$\hbar\rightarrow 0$ the results obtained within TDVP become exact
(to second order in $\hbar$), meaning that one obtains the classical
description of the system dynamics. 

Both in the exact case, and in the TDVP approximation, the role of 
phase $\varphi(t)$ given by (\ref{tdp}) is particularly simple
at the fixed points of (\ref{caneq}). In fact in this case
$\vert \Phi(t)>$ is independent of $t$, and the second term at the r.h.s.
of (\ref{tdp}) (the so called kinetic term) is vanishing. This implies
that $\vert \Psi(t)> =\exp{(-\dsp{i\over\hbar} {\cal H} t)} \vert \Phi_0>$ is
uniquely determined by the energy pertaining 
to the initial state $\vert \Phi_0>$. Such behavior is very reminiscent
of what is called the dynamical phase for a time-dependent
Hamiltonian in the adiabatic approximation. More generally, by
inserting (\ref{tdp}) into the expression (\ref{psi1}) for the state
$\vert \Psi(t)>$, the latter can be written as
\be
\vert \Psi(t)>={\rm exp}\dsp{ \left[ {i\over\hbar} \int dt (-{\cal H}+i\hbar
<\Phi(t)\vert {\partial\over{\partial t}}\vert \Phi(t)>)\right ]}\vert \Phi(t)> 
\quad , \label{psi}
\ee
and we recognize a formal analogy between the phase $\varphi(t)$ and the
dynamical plus geometric phase in the adiabatic approximation
(for a derivation of the quantum phase beyond the adiabatic
approximation, see for instance ref. [10]).
More precisely, in (\ref{psi}) we can identify the first term in
the exponential with the dynamical phase, and the other (kinetic) term
as the geometric phase obtained by relaxing some of the hypotesys
of the adiabatic approximation.
We recall that the latter is nothing but the so called Berry\cite{BER}
phase.
In fact in some simple exactly solvable case\cite{FUTS} it was shown that
--by imposing appropriate quantization condition-- the phase (\ref{tdp})
does coincide with the geometric plus dynamical phases even if evaluated
within the generalized TDVP approximation scheme. This leads us to expect 
that the present approach, apart from leading to a simplified if approximate
description of the dynamics of the Hubbard model, may give a precise physical
information, {\sl i.e.} which is the Berry phase of the states
we are studying. If this is the case, we expect that whenever
we shall obtain states with the same energy ${\cal H}$ but different
phases $\varphi(t)$, an appropriate physical device should be able to
observe their interference.  

A further investigation of this relationship for the model discussed here
is beyond the purpose of the present paper. Here we want just to emphasize
that the state $\vert \Psi(t)>$, thanks to its phase $\varphi(t)$, is able in
principle to approximate the wave-function of the full model
Hamiltonian, no matter how small is the subalgebra ${\cal A}$ of
${\cal G}$. However, we expect in general that the results
will be the more reliable the more ${\cal A}$ is a reasonable description of
${\cal G}$, {\sl i.e.} the more the Hamiltonian is in a (thermodynamical)
phase in which the relevant operators are contained in ${\cal A}$.  
It will turn out that precisely $\varphi (t)$ will measure
how far the system is from the states generated by ${\cal A}$.
In particular, we expect that the system is correctly described by
${\cal A}$ whenever $\varphi(t)$ happens to be linearly increasing
with time, in that the wave-function $\vert \Psi (t)>$ which satisfies 
(\ref{schr}) results to differ by the one we constructed
in ${\cal A}$ ($\vert \Phi(t)>$) just for an oscillating phase factor.

\section{Semiclassical equations of motion in SC phase}

The model described by (\ref{hamhub}), has been intensively studied
in the literature\cite{MON}. However, as for most many-electron problems,
it is quite difficult to obtain rigorous results (for a recent
review, see \cite{LIE}). In particular, only the one-dimensional
zero-temperature energy is known exactly\cite{LIWU}. Therefore,
different approximation schemes have been used in order to deal with
(\ref{hamhub}). Among them, a standard one is the Hartree-Fock
decoupling procedure, which amounts in approximating the interaction
term by reducing it to a sum of its bilinear parts, weighted by
coefficients which have to be fixed self-consistently. There are of
course different ways of decoupling the interaction term, depending
on the phase which has to be investigated \cite{HIR}, \cite{LIMO}.
The dimension of the dynamical algebra of the resulting
"decoupled" Hamiltonian
turns out to be greatly reduced, and to all effects one is lead to
deal with a subalgebra of the spectrum generating algebra of the original
Hamiltonian. It is our purpose to construct the GCS involved by
TDVP scheme in these subalgebras.

In order to identify the subalgebra, we explicitly need the reduced
Hamiltonian.  Neglecting for the moment the possibility of
an AF or a CDW phases (which will be investigated in a later section)
and of a ferromagnetic phase (which is not to be expected at low $U$),
it turns out that such Hamiltonian coincides with the linearized  
Hamiltonian $H_l^{(sc)}=\sum_k H_k^{(sc)}$, where
\bq
H_k^{(sc)} = (\epsilon_k - \mu) n_k +U\left [{n\over 2} n_k + 
(\Delta a_k^\dagger a_{-k}^\dagger + h.c.)\right ]-U ( {n^2\over 4} N
+ \vert \Delta\vert ^2)N \quad . \nonumber 
\eq
Here, as customary, $k$ stays for the multi-index $({\bf k}, \sigma)$
($ k\equiv ({\bf k},\uparrow)$ , $-k\equiv (-{\bf k},\downarrow)$), 
and $n_k\doteq a_k^\dagger a_k$. Moreover $n$ is the average
electron number per site, $n=\dsp{1\over N} <{\cal N}_e>$, and $\Delta
\doteq \dsp{1\over N} <\sum_k a_{-k} a_k>$ is the average pairing per
site, where $<\bullet>$ denotes the expectation value of operator
$\bullet$ over appropriate states.

It is important noticing that $H_l^{(sc)}$, contrary to $H_{Hub}$, for any
$\Delta\neq 0$ does not commute with the electron-number operator
per-site ${\cal N}_e$. This is justified by observing that
$H_l^{(sc)}$ is a faithful approximation of $H_{Hub}$ in an ordered phase
which does not conserve such quantity (namely, the superconducting phase). 
On the other hand for $\Delta=0$ ${\cal N}_e$ is still conserved, and
$H_l^{(sc)}$ describes in that case the system in a paramagnetic phase,
which is known to be the case for the ground state, at least for low positive
$U$, away from half filling ({\sl i.e.} $n\neq 1$).

The Hamiltonians $H_k^{(sc)}$'s have the property that 
$[H_k^{(sc)}, H_{k'}^{(sc)}]=0$, and hence can be diagonalized simultaneously.
More precisely, $H_l^{(sc)}$ turns out to be an element of the dynamical
algebra
${\cal A}_{sc}= \bigoplus_k{\cal A}_k^{(sc)}$, where
${\cal A}_k^{(sc)}$ is the local $su(2)$ generated by
\be
{\cal A}_k^{(sc)} = \left\{J_3^{(k)}\equiv{1\over 2}
(n_k+n_{-k}-1),J_+^{(k)}\equiv
a_k^\dagger a_{-k}^\dagger,J_-^{(k)}\equiv a_{-k} a_k\right\} \quad .
\label{algsc}
\ee
Any eigenstate of the Hamiltonian $H_l^{(sc)}$ can then be
expressed as superposition
of the GCSs of ${\cal A}_{sc}$, namely 
\bq
\vert \eta> =\Pi_k (1+\bar\eta_k \eta_k)^{-{1\over 2}}\exp{\left (\eta_k
J_+^{(k)}\right )}\vert 0>_{sc} \quad , \nonumber
\eq
where the coefficients $\eta_k\in {\Bbb C}$ parametrize the (overcomplete) set
of exact GCS of $H_k^{(sc)}$, and $\vert 0>_{sc}$ is the electron vacuum.

In line with the general approach discussed in the previous section we
can think of the $\eta_k$'s as time dependent parameters, and
construct the approximate trial time dependent wave-function of the full
Hamiltonian $H_{Hub}$, $\vert \psi(t)>_{sc}$, as
\be
\vert \psi (t)>_{sc}= {\rm e}^{{i\over\hbar} \varphi_{sc}(t)}\vert \eta (t)>
\quad .\label{psisc}
\ee

We are now ready for evaluating the expectation value of $H_{Hub}$ over
$\vert \psi(t)>_{sc}$, namely the semi-classical Hamiltonian ${\cal H}_{sc}$,
which reads  
\be
{\cal H}_{sc} = 2\sum_k (\epsilon_k-\mu) {\bar\eta_k \eta_k \over{1+
\bar\eta_k\eta_k}}+ U \left
[\left ( \sum_k {\bar\eta_k \eta_k\over{1+\bar\eta_k \eta_k}}\right
 )^2+\sum_{k,l}{\bar\eta_k \eta_l \over{(1+\bar\eta_k \eta_k)
(1+\bar\eta_l \eta_l)}}\right ] \quad , \label{hamsc}
\ee
where $\eta_k$, $\bar\eta_k$ obey the Poisson-braket relations obtained from
(\ref{pb}), $i \hbar \{\eta_k,\bar\eta_k\}=(1+\bar\eta_k \eta_k)^2$.

Instead of proceeding directly to the derivation of the canonical
equations of motion, we notice that Hamiltonian (\ref{hamsc}) can
be fruitfully rewritten in terms of the following semiclassical
pseudospin variables
\be
S_3^{(k)}\doteq{1\over 2} \frac{\bar\eta_k\eta_k-1}{\bar\eta_k\eta_k+1}
\equiv_{sc}<\phi(t)\vert J_3^{(k)}\vert \phi(t)>_{sc}
\quad ,\quad S_+^{(k)}\doteq   \frac{\bar\eta_k}{1+\bar\eta_k\eta_k}
\equiv_{sc}<\phi(t)\vert J_+^{(k)}\vert \phi(t)>_{sc} \quad ,\label{spin}
\ee
and $S_-^{(k)}=(S_+^{(k)})^*$, whose Poisson brackets recover for each
$k$ a $su(2)_k$ algebra. Explicitly
\be
i \hbar \left \{S_+^{(k)},S_-^{(k)}\right\}= 2 S_3^{(k)}\quad ,\quad
i \hbar \left \{S_\pm^{(k)},S_3^{(k)}\right\}= \mp S_\pm^{(k)}
\quad ,\label{su2}
\ee
Moreover, one can define the related "mesoscopic" variables
$S_\alpha^{(a)}\doteq\sum_{k\in\tilde\Lambda_a} S_\alpha^{(k)}$,
with $\alpha=3,+,-$, $\tilde\Lambda_a\doteq\{k\in\tilde\Lambda;\epsilon_k
=\epsilon_a\}$ denoting the mesoscopic (kinetic energy) levels.
One can easily verify that the
$S_\alpha^{(a)}$'s form a $su(2)$ algebra like (\ref{su2}) (with
$k\rightarrow a$), which we identify by $su(2)_a$.
Hamiltonian (\ref{hamsc}), when written in terms of $S_\alpha^{(a)}$,
reduces to a genuine one-dimensional problem, in that the index $a$ 
(contrary to $k$) is strictly one-dimensional,
numbering the different mesoscopic levels.
Indeed
\be
{\cal H}_{sc}=2\sum_a (\epsilon_a -\mu) (S_3^{(a)}+{N_a\over 2})+u\left[\left
(S_3+{N\over 2}\right )^2+\left\vert S_+\right\vert ^2\right] \quad .
\label{hamspinsc}
\ee
Here $S_\alpha \doteq \sum_a S_\alpha^{(a)}\equiv\sum_k S_\alpha^{(k)}$, 
and $u=U/N$. The Casimir operators of both 
$su(2)_k$ and $su(2)_a$ algebras, $I_k\doteq
\vert  S_+^{(k)}\vert ^2+\vert S_3^{(k)}\vert ^2$ (and the same definition for
$I_a$ with $k$ replaced by $a$) are conserved quantities for
${\cal H}_{sc}$. In view of the definitions (\ref{spin}),
$\dsp{I_k= {1\over 4}}$, and $\dsp{I_a\leq {N_a^2\over 4}}$,
depending on the initial conditions.

Noticeably, Hamiltonian (\ref{hamspinsc}), like $H$ and unlike $H_l^{(sc)}$,
commutes also with $S_3$, {\sl i.e.} the semi-classical variable
corresponding to the total electron number operator. $S_3$ is thus, as
it should be, a conserved quantity, for which the relation holds  
\be
S_3 = {N\over 2}(n-1) \quad . \label{s3}
\ee
In this sense we can therefore claim that the $1-d$ Hamiltonian ${\cal H}_{sc}$
obtained by means of the present semi-classical approach
is a more accurate approximation of $H$ than $H_l^{(sc)}$.
In particular, in (\ref{hamspinsc}) the $k$-modes are coupled
dynamically through $\vert S_+\vert ^2$, while in ${\cal H}_l^{(sc)}$
they are not. This feature  in turn keeps track in the present scheme
of the non linearity of $H_{Hub}$, thus making ${\cal H}_{sc}$
a good candidate for giving an approximate description of the physics
of the Hubbard model in the whole phase-space. Of course, 
the results obtained by using instead $H_l^{(sc)}$ will be reproduced by the
present approximation, as we shall see in the next section. 

From (\ref{su2})-(\ref{hamspinsc}) we can now derive the equations of
motion for the mesoscopic variables $S_\alpha^{(a)}$, which read
\bq
\mp i\hbar \dot S_\pm^{(a)} &=& \delta_a S_\pm^{(a)} 
- 2 u S_3^{(a)} S_\pm \nonumber\\
i \hbar \dot S_3^{(a)}&=& u\left [S_- S_+^{(a)}-
S_+ S_-^{(a)}\right ] \quad . \label{eqmotsc}
\eq
Here $\delta_a=2(\epsilon_a-\mu)+u n N$, where the constant factor
$u n N-2\mu$ in $\delta_a$ is vanishing at half-filling
($\mu={U\over 2}$, $n=1$), and in any case does not affect the dynamics
described by $S_\pm^{(a)}$, apart from an over-all phase factor
e$^{\pm{i\over\hbar}(U n-2\mu)\tau}$. Notice that of the three
equations (\ref{eqmotsc}) only two are independent, whereas the
third one is obtained from the Casimir constraint. For instance,
one could use as independent variables $S_3^{(a)}$, which fixes
also the absolute value of $S_\pm^{(a)}$, and the phase $\lambda_a$ 
of $S_\pm^{(a)}=\vert S_\pm^{(a)}\vert  {\rm e}^{\pm i\lambda_a}$. This
alternative representation of the pseudo-spin variables will also be considered,
when useful, in the text.

Let us emphasize that the true dynamical variables are
of course the microscopic canonical variables $S_\alpha^{(k)}$,
which satisfy the same equations of motion (\ref{eqmotsc}) with $a$
replaced by $k$. Here we preferred to write them only for the
mesoscopic variables $S_\alpha^{(a)}$ because the Hamiltonian
${\cal H}_{sc}$ given by (\ref{hamsc}) was shown to be degenerate
with respect to the inner dynamics of the mesoscopic variables.
Moreover every solution we will be able to find for the $S_\pm^{(a)}$'s
holds straightforwardly also for the $S_\pm^{(k)}$'s, as
$\delta_k\equiv\delta_a$. In fact, apart from this simple
case, every solution for the microscopic variables can be
in principle worked out once we have found the mesoscopic solutions,
and consequently $S_\pm$, as in this case the equations (\ref{eqmotsc})
with $a\rightarrow{k}$ reduce to a linear system with time-dependent
coefficients. Interestingly, it is
easily verified from (\ref{eqmotsc}) that the scalar
product of any two microscopic pseudospin vectors ${\bf S}^{(k)}$
belonging to the same mesoscopic level is constant. This observation
implies that in fact the time evolution of every microscopic vector
in a given mesoscopic level is identical, the relative orientation 
of different ${\bf S}^{(k)} (t)$ depending only on the initial
conditions.

According to the generalized TDVP approach introduced in previous
section, and by means of (\ref{tdp}), (\ref{spin}), (\ref{hamspinsc}),
and (\ref{eqmotsc}), we are finally able to obtain the
time-derivative of the time-dependent phase $\varphi_{sc}(t)$,
\bq
\dot \varphi_{sc} &=& - {\cal H}_{sc}+ i\hbar\sum_k
\frac{\dot S_-^{(k)} S_+^{(k)}
-\dot S_+^{(k)} S_-^{(k)}}{1-2 S_3^{(k)}}\nonumber\\
&=& u\left\{n^2{N^2\over 4} -\left [S_+\sum_k S_-^{(k)}
\frac{1+2 S_3^{(k)}}{1-2 S_3^{(k)}}+c.c.\right]\right\}\nonumber\\
&=& -{\cal H}_{sc}+{1\over 2}\hbar\sum_k(1+2 S_3^{(k)})\dot
\lambda_k \quad , \label{phasesc}
\eq
where the last expression was explicitly written to
make evident that a non-vanishing geometric contribution to
$\varphi_{sc}$ is expected whenever the phase $\lambda_k$ of
$S_\pm^{(k)}$ is not constant.

Eq. (\ref{phasesc}) has some other relevant features which it is
worth underlying:
\begin{itemize}
\item{} it vanishes for vanishing $u$, as can be recognized from the
second line form. In fact we know that if
this is the case the wave-function given by (\ref{psisc}) becomes exact, and
according to the discussion developed in previous section this
implies that $\varphi_{sc}(t)$ must reduce to the exact value
given by (\ref{tdp}), which can be shown to be zero;

\item{} it reduces to the constant $u n^2 \dsp{N^2\over 4}$ for $S_+=0$.
Being $S_+$ related through (\ref{spin}) to the semi-classical analog
of the total pairing operator, it must be inferred that a non-linear
time behavior of $\varphi_{sc}(t)$ is closely connected to the possible
superconductivity of the state;

\item{} contrary to both ${\cal H}_{sc}$ and the equations of motion
(\ref{eqmotsc}), $\varphi_{sc}(t)$ can not be expressed only in terms 
of the one-dimensional mesoscopic variables $S_\alpha^{(a)}$.
This means that it maintains memory of the inherent complexity of the
original Hamiltonian, and gives information about the time-evolution of
its wave-function which goes beyond that implicit in its semi-classical
approximation (\ref{hamspinsc}), in particular depending on the
inner dynamics of the mesoscopic levels $a$.

\end{itemize}

\section{Fixed points and stationary points}

The first step in investigating the dynamical behavior of any nonlinear
Hamiltonian system usually consists in finding its fixed points, that is
those points in phase space where the equation of motions involve vanishing
time derivatives of the dynamical variables. The stability analysis of such set
of points leads to revealing their topological nature (by resorting, for example,
to standard methods such as Routh-Hurwitz criterion) and, in conclusion,
to structuring the phase space in regions where the dynamical behavior
of the system exhibits well defined features \cite{ARPL}.

A complete stability analysis is beyond the scope of present work.
In fact, in this section we shall simply work out all the solutions
to fixed point equations, in particular showing that indeed those among 
them which minimize the energy ${\cal H}_{sc}$ give the same
energy and the same self-consistency equation as the Hartree-Fock
approximation, both for $u\leq 0$ and for $u\geq 0$. Apart from that,
the knowledge of the fixed points allows in principle to look for
other solutions of (\ref{eqmotsc}) by means of standard perturbative 
methods in their proximity. 

Minimum energy points are contained among the stationary points of
${\cal H}_{sc}$, which are easily shown to coincide with fixed points
of (\ref{eqmotsc}) first by rewriting the equations of motions in terms
of canonical variables $( S_3^{(a)},\lambda_a )$, and by setting then 
$\dot {\lambda_a} = 0, \dot S_3^{(a)} =0 $. Since this is equivalent
to $\dot S_{+}^{(a)} = 0$ and $\dot S_{3}^{(a)} = 0$,
eqs. (\ref{eqmotsc}) furnish the stationary point equations
\bq
0 &=& \delta_a S_{+}^{(a)} - 2 u S_{3}^{(a)} S_{+} \quad, 
\label{fp1} \\
0 &=& S_{+}^{(a)} S_{-} - S_{-}^{(a)} S_{+} \quad.
\nonumber 
\eq
The case in which $S_{+}^{(a)} = 0$ for any $a$ represents the simplest
possible solution. We observe that $S_{3}^{(a)}$, thanks to the Casimir's
constraint,  can be chosen in a fully
arbitrary way within the interval $\dsp{-{N_a\over 2}\leq S_3^{(a)}\leq
{N_a\over 2}}$, so that an enormous number of stationary points characterizes
the mesoscopic pseudospin dynamics.

It is important noticing that the solution $S_\pm^{(a)}=0$, when
inserted in the equations for the microscopic variables, makes them
immediately integrable, and the solution shows that in general a
microscopic, inner dynamics for $k$-pseudospins $\in \tilde\Lambda_a$
can take place, according to
\be 
S_{+}^{(k)}(t) = S_{+}^{(k)}(0) \,\, e^{i t {\delta_a\over \hbar}}
\quad ,\label{topo}
\ee
provided $\sum_{k \in \tilde\Lambda_a} S_{+}^{(k)}(0)\doteq
\sum_{k \in \tilde\Lambda_a} R e^{i \lambda_k(0)} = 0$. Such a constraint
in $d=2$ is naturally obeyed by those configurations where the initial  
phase $\lambda_k(0)$ is topologically nontrivial while $R$ is independent
of $k$. Indeed this is the case when $\lambda_k(0)$ --regarded as a
function of $k$ along the 1-$d$ closed paths associated to each
$a$-th mesoscopic level-- undergoes a variation
of $2 \pi p$, with $p\in{\Bbb N}$. For paths with energy $\epsilon_a
\simeq 0$ the number of modes $N_a$ is great enough to allow
$e^{i \lambda_k(0)}$ to be twisted many times in a quasi continuous way.
Let us underline that solution (\ref{topo}), which --being consistent
with $S_\pm^{(a)}=0$-- corresponds to a stationary point of ${\cal H}_{sc}$,
is not a fixed point of the microscopic dynamics
when $S_\pm^{(k)} (0) \neq 0$. 

The energy associated with the solution $S_+ = 0$ has the form
$E = - u(S_3 + N/2)^2 + \sum_{a} \delta_a (S_3^{(a)}+{N_a\over 2})$.
For $u > 0$ it is easy to check that an absolute minimum endowed with
the energy
\be
E_{sc}^{(+)} = - u n^2 {N^2\over 4}- \sum_{a>F} \vert 
\delta_a\vert  N_a\label{gs}
\ee
is reached when $S_3^{(a)} = +(-) \dsp{N_a\over 2}$ for $a>F \,(a < F)$
is imposed, $F$ being that particular value of $a$ for which
$\delta_{F}=0$, which implies $\mu=\epsilon_{F}+U \dsp{n\over 2}$,
$\sum_{a>F}N_a-\sum_{a<F}N_a=N(n-1) - 2 S_3^{(F)}$, and $\delta_a=
2(\epsilon_a- \epsilon_F)$.
This absolute minimum corresponds to $\vert S_\pm^{({k})}\vert =0$ for each
${k}$. Noticeably, the latter constraint does allow
the ground state to still have a phase dynamics. In fact, on the one
hand, by rewriting the equations of motions in terms of the canonical variables
$\lambda_a, S_3^{(a)}$ introduced in previous section, it is
straightforwardly verified that in the limit where
$\vert S_\pm^{(k)}\vert \rightarrow 0$ uniformly the equation for the angle
variables reduces to
\be
\hbar \dot\lambda_k =\delta_k- 2 u S_3^{(k)}\sum_{\elle\in\tilde\Lambda}
\cos(\lambda_k-\lambda_\elle) \quad , \label{eqlambda}
\ee
where, to second order in $\vert S_\pm^{(k)}\vert $,
$S_3^{(k)}=\pm{1\over 2}$ for $k\in\tilde\Lambda_\mp$. 
On the other hand for $\vert S_\pm^{(k)}\vert \equiv 0$ the phase is totally
free, as one can check from equations (\ref{fp1}). Hence, by a continuity
argument, we expect that also in this case the dynamics of $\lambda_k(t)$
evolves according to (\ref{eqlambda}). Equation (\ref{eqlambda}) has already
been investigated in a different context (see for instance \cite{SAKU}). In
particular, it was shown \cite{KU} that for $XY$-like models it allows for
vortex-like excitations. 
Moreover, in the continuum limit it can be recognized as a Bernoulli-like
equation, the latter being known to describe once more a vortex dynamics.
Finally, let us observe that the solution of (\ref{eqlambda}) contains
as a particular case the (topological) one discussed after eq. (\ref{topo}),
which requires $S_\pm^{(k)}\neq 0$, and reduces to it only in the (exact)
non-interacting case, {\sl i.e.} for $u=0$. 

The absolute minumum (\ref{gs}) corresponds to the paramagnetic phase
within the Hartree-Fock approximation, and gives the same ground state
energy. Contrary to that approximation,
here it was possible to make evident a non-trivial dynamical behavior
of the paramagnetic ground state. Such behavior implies in
particular the appearance of a non-vanishing geometric phase in the
ground state, as can be understood from the third of the
equations (\ref{tdp}). Let us recall that this should happen
at any filling but half. We shall see in fact that
at half-filling ($n=1$) states built with antiferromagnetic
order can provide lower energy for the corresponding semiclassical
Hamiltonian, again in agreement with the Hartree-Fock approximation. 

Moreover, let us stress that stationary points characterized by
$S_+^{(a)} = 0$ --even when not identifying an absolute minimum--
indeed can be shown to be local minima of the Hamiltonian when the
geometric constraints represented by the Casimir's are taken
into account. A simple first order expansion of ${\cal H}_{sc}$ in
the variables
$\vert  S^{(a)}_{+} \vert ^2$, where $S_{3}^{(a)}$'s are now expressed as
$S_{3}^{(a)} = +(-)(I_a - \vert  S^{(a)}_{+} \vert ^2)^{-1/2}$ for $a>F
(a<F)$, shows that the variation $\delta {\cal H}_{sc}$ is positive
provided $u$ is positive and sufficiently small. In summary, we conclude
that such stationary points are minimum energy points for $u>0$, possibly
possess inner dynamics and topological structure, but do not involve
superconductive situations, being $S_+ = 0$.

The remaining set of fixed points -- which are still solutions of   
equations (\ref{fp1}) -- can be fruitfully parametrized
through the parameters $I_a$ and $S_+$. Explicitly
\be
S_3^{(a)}=-s_a\delta_a\sqrt{I_a\over{\delta_a^2+4 u^2 \vert S_+\vert ^2}}
\quad ,\quad  S_+^{(a)} = 2 s_a \vert u\vert \sqrt{I_a\over{\delta_a^2+4 u^2
\vert S_+\vert ^2}}\vert  S_+\vert 
\quad , \label{fixedsc}
\ee
with $s_a=\pm 1$. $S_+$ does not play the role of a free
parameter, but it turns out to be constrained by the equation 
\be
1 = - 2 u \sum_a s_a \sqrt{I_a\over{\delta_a^2+4 u^2 \vert S_+\vert ^2}}  
\quad . \label{scsc}
\ee
By substituting (\ref{fixedsc}) in (\ref{hamspinsc}), and choosing the
values of $s_a$ and $I_a$ which minimize ${\cal H}_{sc}$ ($s_a=1$ and
$I_a=\dsp{N_a^2\over 4}$), it is seen from (\ref{scsc}) that such solution
exists only for $u<0$, and corresponds to an energy
\be
E_{sc}^{(-)} =  \vert u\vert  {N^2\over4} n (2-n) -{1\over 2}
\sum_a N_a \sqrt{\delta_a^2+4 u^2 \vert S_+\vert ^2} + \vert u\vert  \vert S_+
\vert ^2 \quad . \label{emsc}
\ee
This result is once more in agreement with the ground state result
for the superconducting regime within the Hartree-Fock approximation,
as can be seen by identifying the variational parameter $\Delta$
with the semiclassical pairing operator $S_+$. In particular, 
the constraint equation (\ref{scsc}) coincides with the self-consistency
equation for $\Delta$.

Let us notice that --as opposite to the mesoscopic fixed points
$S_\pm^{(a)}=0$ case-- here the insertion of the solutions
(\ref{fixedsc}) into the equations of motion for the 
variables $S_\pm^{(k)}$'s does not allow any microscopic
dynamics, as the constraint $S_\pm^{(a)}\neq 0$ has to be satisfied.

\section{Collective frequency dynamics}

The dynamical system described by semiclassical eqs. (\ref{eqmotsc}) 
is integrable in the special case when the pseudospin variables $S^{(a)}_3$ are 
supposed to be time-independent. The main effect of such an assumption
is, in fact, of halving the number of the system degrees of freedom.
This can be easily seen by
observing that $\vert  S^{(a)}_{+} \vert $ cannot depend on time
consistently with the fact that the Casimir's 
$I_a$ are constants of motion, so that only the phases of the pseudospin
projection variables $S_{+}^{(a)}$ are allowed to depend on time.
Further restrictions on the dynamics are due to the
equations of motion which take the form
\bq
-i \hbar \dot S_{+}^{(a)} &=& \delta_a S_{+}^{(a)} - 2 u S_{3}^{(a)} S_{+} 
\quad ,\label{de3} \\
S_{+}^{(a)} S_{-} &=& S_{-}^{(a)} S_{+} \quad.
\nonumber
\eq
The first of eqs. (\ref{de3}) show how the system formally reduces
to an ensemble of interacting oscillators with coupling constants
$2 u S_{3}^{(a)}$. Moreover, together with the second, rewritten as 
\be
S_{+}^{(a)} / S_{-}^{(a)} =  S_{+} / S_{-}   \quad ,
\label{de5} 
\ee
state that a unique, time-dependent phase $\dsp{W\over\hbar}\tau$
characterizes the system dynamics. Namely, for any $a$,
\be
S_{+}^{(a)}(\tau) = V_a e^{i \alpha} e^{i {W\over\hbar} \tau}\quad ,
\label{sol}
\ee
\noindent with $V_a, \alpha, W\in {\Bbb R}$. Both the constant phase $\alpha$
and $V_a$ are fixed by assigning the initial conditions $S_{+}^{(a)}(0)$.
The linear character of eqs. (\ref{de3}) allows one to recast them in the
matrix form
\be
({\bf M} \{ {\bf {\cal S}} \}- W{\Bbb I} )\cdot {\bf S_+}= 0 \quad , 
\label{me} 
\ee   
where the vector ${\bf S_+}$ has components $S_{+}^{(a)}$ and the dynamical
matrix ${\bf M}$, whose elements can be obtained by system (\ref{de3}),
explicitly depends on the set ${\bf {\cal S}} = \{ S_3^{(a)} \}$.
The associated secular equation, which in turn provides the eigenvalue
equation, 
\be
det( {\bf M} - W {\Bbb I})\,\,=
\left [ \quad 1 + 2u \sum_b { \frac {S_3^{(b)}} {W - \delta_b}}  \quad \right ] 
\Pi_a (W - \delta_a) = 0 \quad ,
\label{det} 
\ee
is polynomial in $W$. 
The eigenvector components $V_a$ can now be expressed
in terms of $W$, $V = \sum_a V_a$ and $S_3^{(a)}$ as
\be
V_a =
{\frac {2u S_3^{(a)}} {\delta_a - W}} V \quad. \label{va}
\ee 
It should be noticed that in fact $V$ is itself a function of the
initial conditions $S_3^{(a)}$ and energy ${\cal H}_{sc}$, through the
relations (\ref{hamspinsc}) and (\ref{sol}), which give 
$V = \pm [( {\cal H}_{sc}+ u n^2 N^2/ 4 - 2t\,
\sum_a \delta_a S_3^{(a)}) / u ]^{1/2}$.

Moreover the eigenvalues fulfilling eq. (\ref{det}) are obtained,
after assigning the initial condition set $\{S_3^{(a)} \}$, by
solving  
\be
1 = 2u \sum_b { \frac {S_3^{(b)}} {\delta_b - W}}  \quad .
\label{a1}
\ee
It turns out that the factor $\Pi_a (W - \delta_a)$ in (\ref{det}) does not play
a role unless $S_3^{(a)} = 0$ for some $a$. When this is
the case some of the eigenvalues coincide with the system proper frequency
$\delta_a$. The total number of eigenvalues, corresponding
to the number of different mesoscopic levels,  is however kept
constant.

Both eigenvalues and eigenvectors can be easily obtained in an
(approximate) explicit way when $\vert  u \vert  /t$ is suitably small.
Looking at the structure of (\ref{a1}), it appears clear that the values of
$W$ close enough to $\delta_a$ are reasonably expected to fulfill it. In order
to check this we first replace $W$ with $W_a =\varepsilon + \delta_a$ in
eq. (\ref{a1}) which becomes
\bq
1 =  - { \frac { 2u S_3^{(a)}} {\varepsilon}} +
2u \sum_{b \ne a} { \frac {S_3^{(b)}} {\delta_b - \delta_a - \varepsilon}}
\quad, \nonumber
\eq
then, by taking
$\vert  \varepsilon \vert  \ll \vert  \delta_b - \delta_a \vert $, for any pair
$(a,b)$, one easily finds that
$\varepsilon \simeq - 2u S_3^{(a)}$ thus obtaining
\bq
W_a \simeq \delta_a - 2u S_3^{(a)} \quad. \nonumber
\eq
\noindent On the other hand,
the condition $\vert  \varepsilon \vert  \ll \vert  \delta_b - \delta_a \vert $
is satisfied if it holds in the less favourable case $a=0 , b=\pm 1$.
Since 
$\vert  \delta_0 - \delta_1 \vert  = 2t \epsilon_1 \simeq 8t \pi^2 / N$
(for $d=2$), then
the condition on $\varepsilon$ becomes $\vert  \varepsilon \vert  \ll 8t \pi^2 /
N$, which finally leads to 
\bq
\vert  U \vert  \ll 8t \pi^2 / N_a
\nonumber
\eq
\noindent for the greatest possible $S_3^{(a)}$ given by
$\dsp{N_a\over 2}$. The present approximation scheme, based on
considering $W_a \simeq \delta_a$, is thus permitted for reasonably small
values of $\vert  U \vert  /t$. In particular, if $\vert U\vert \ll8 t
\pi^2/N_0$ all the eigenvalues can be obtained from this scheme, whereas if
$\vert U\vert \geq 8 t \pi^2$ none of the $W_a$'s  is well approximated by it. 
The $a$-th eigenvector associated with eigenvalue $W_a$ is readily obtained
from eqs. (\ref{va}) and exhibits components $V_c(W_a)$ given by
\bq
V_c(W_a) \simeq {\frac {2u S_3^{(c)}} {\delta_c - \delta_a}} \,\, V
\quad for \,c\neq a \quad, \quad
V_a(W_a) \simeq
V \,\,
\left (1 -{\sum_{b \ne a} { \frac {2u S_3^{(b)}} {\delta_b - \delta_a} }}
\right) \quad.
\nonumber
\eq
The above equations show how the $a$-th eigenvector ${\bf S_+} (W_a)$
in this approximate case is characterized by the fact that only the component
$V_b$ with $b = a$ is strongly nonzero, being in fact $V_a \simeq V$.
This implies that each eigenvector can be
regarded as describing a superconductive situation where the superconductive
order parameter $ S_+ = < \sum_k J_+^{(k)} > = \sum_k S_+^{(k)} $ is
essentially given by $S_+ \simeq  S_+^{(a)}$ and the $u$-dependent contribution
to the energy ${\cal H}_{sc}$ is mainly given by the $k$-modes with
$\epsilon_k = \epsilon_a$.

Some general observations are now in order. First we notice that the eigenvector
problem is completely solved provided $\{ S_3^{(a)} \}$ and
${\cal H}_{sc}$ --the quantities which, at this stage, describe the initial
system configuration-- have been assigned, and the eigenvalues $W_a$ have
been worked out from (\ref{a1}). No restriction constrains ${\cal H}_{sc}$
and $\{ S_3^{(a)} \}$ except for the filling condition (\ref{s3}) and the
condition $V\neq 0$. The latter allows one to consider eigenvectors with
arbitrarily small components $V_a$ but excludes the solutions characterized
by $\bf S_+ = 0$ ($V_a = 0$ for any $a$) representing a subset of the
solution of the fixed point equation (\ref{fp1}). The single-mode solution
set is thus completely disjoint from such fixed point subset in the space
of solutions of eqs. (\ref{de3}) even if the former is dense
around any element of the latter. On the contrary, the other fixed
points of (\ref{de3}), given by (\ref{fixedsc}), are a (time-independent) 
subset of (\ref{sol}), corresponding to $W=0$. 

Moreover we point out that the nonlinear nature of pseudospin dynamics
survives our initial assumption $S_{3}^{(a)} = const$ because of the
second of eqs. (\ref{de3}). In fact the linear system of coupled
oscillators described by eqs. (\ref{de3}) should have an arbitrary
superposition of eigenvectors related to eq. (\ref{me}) as a general solution.
This is no longer possible when eqs. (\ref{de5}) are taken
into account in that any superposition of single-mode solutions (eigenvectors)
violates the request that pseudospins exhibit the same phase. 

Furthermore we observe that
$S_+^{(k)}(\tau)$ can be easily obtained from eqs. (\ref{de3}),
where the term $S_+(\tau)$ is now playing the role of an external forcing
term. Since $S_+^{(k)}(\tau)$ results to be proportional to $S_+^{(a)}$ up
to a constant factor $e^{i \theta}$, then it follows that the equation for
quantum phase $\varphi_{sc}(\tau)$ has form
\be
\hbar {\dot \varphi}_{sc}(\tau) = W ( S_3 + N/2) - {\cal H}_{sc} \quad.
\ee
\noindent
Hence, already in this simple integrable case within our approximation
there is a non-vanishing contribution of the geometric phase (equal
to $W n N/2 $ times $t$), at any energy but the ground state. Such contribution
should in principle be observable by appropriate experiments.
  
Finally it is remarkable that these single-mode
solutions, exhibiting some form of collective order through the unique
time-dependent phase ${W\over\hbar} t$, correspond to non-vanishing
superconductive order parameter $S_+$.

\section{Slow dynamics vs fast dynamics} 
\bigskip
A standard procedure for tackling many-body system dynamics consists in
simplifying the equation of motions by separating fast degrees of freedom
from slow degrees of freedom \cite{FUTS,SHWI,ME}.
Such a procedure is profitable in that it
leads the slow variable-system to become an autonomous system and sometimes
reduces the complexity of its equations of motion. These features are, of
course, appealing here because the slow variables dynamics is the one surviving
at a macroscopic level (and thus it might be observable), while dynamics of
high frequency degrees of freedom disappears on large time scales.

For the pseudospin system a classification of
pseudospins either as fast variables or as slow variables is naturally
established by the fact that either $\delta_a \simeq 0$ 
or $\delta_a \ne 0$ respectively. We recall that such reference
parameter depends on $n$ and is associated
with the mesoscopic $a$-level of the ground state configuration where
$S_3^{(a)}$ changes its sign.

The effect of such a distinction is made evident by performing the
substitutions $S_{+}^{(a)}=exp(i \delta_a {t\over \hbar} ) \Psi_a$,
which turn eqs. (\ref{eqmotsc}) into the form
\bq
-i \hbar \dot \Psi_a &=& - 2u S_{3}^{(a)} \Psi_a 
- 2 u S_{3}^{(a)} \sum_{b \ne a}
e^{i  (\delta_b - \delta_a){t\over\hbar}} \Psi_b 
\quad, \label{sf1}\\ 
i \hbar \dot S_{3}^{(a)} &=& u 
\Bigl ( \Psi_a \sum_{b \ne a} \Psi_b^{*}
e^{i  (\delta_a - \delta_b){t\over\hbar}}) 
- c.c. \Bigr ) \quad,
\nonumber 
\eq
explicitly exhibiting dependence on the frequencies $\delta_a$.
Introducing the parameter $\delta_*$ as the frequency
distinguishing slow frequencies (defined by
$\vert  \delta_a \vert  \le \delta_*$) from fast frequencies 
(defined by $\vert  \delta_a \vert  > \delta_* $), it clearly
results that those time-dependent oscillating terms of eqs.
(\ref{sf1}) where $\vert  \delta_b \vert  > \delta_*$
can be neglected on a time-scale greater than $ \hbar/ \delta_*$,  
since their rapid oscillations make their time-average vanishing.

This fact has remarkable implications. In fact,
upon denoting fast pseudospins variables and slow pseudospin variables
by $F_{\pm}^{(b)}$, $F_{3}^{(b)}$ and $Q_{\pm}^{(b)}$, $Q_{3}^{(b)}$ 
respectively, we are now able of separating the dynamical equation set
in two almost independent sub-sets, the first one of which describes
short time-interval processes ($t < \hbar/ \delta_*$), and reads
\bq
-i \hbar \dot F_{+}^{(a)} &=& \delta_a F_{+}^{(a)} - 2 u F_{3}^{(a)}
( F_{+} + Q_+ ) \quad , 
\label{sf3} \\
i \hbar \dot F_{3}^{(a)} &=& u \Bigl [ F_{+}^{(a)} (F_- + Q_-) - c.c.
\Bigr ] \quad.
\nonumber 
\eq
while the second concerns long time processes ($t > \hbar/ \delta_*$) 
involving the slow variables, and is given by
\bq
-i \hbar \dot Q_{+}^{(a)} &=& \delta_a Q_{+}^{(a)} - 2 u Q_{3}^{(a)} Q_+ 
\label{sf5} \quad ,\\
i \hbar \dot Q_{3}^{(a)} &=& u ( Q_{+}^{(a)} Q_+ - c.c.)\quad.
\nonumber 
\eq
Here $Q_+={\sum_a}' Q_+^{(a)}$, and $F_+={\sum_a}'' F_+^{(a)}$, where the
prime and the double prime remind that $a$ must range within selected
intervals ($\vert \delta_a\vert >\delta_*$ and $\vert \delta_a\vert <\delta_*$
respectively). We notice that in eqs. (\ref{sf3}) $Q_{\pm}$ can be
regarded as time-independent terms (adiabatic approximation), since their
evolution takes place on the time-scale of slow variables, whereas in eqs.
(\ref{sf5}) fast variables are absent because of the
effects of rapid oscillations discussed above. Also, when
$t > \hbar/ \delta_*$, such oscillations makes $F_{\pm}$ negligibile with
respect to $Q_{\pm}$ in terms like $(Q_{\pm} + F_{\pm})$ of
eqs. (\ref{sf5}), so that $F$-variable dynamics turns out to
be driven by $Q_{\pm}$. On the other hand the $Q$-system
can be considered as an almost isolated system which exhibits the same
features of the initial $N$-pseudospin system except for the fact that now
the pseudospin number is $N_* = {\sum_a}' N_a < N$ and
the effective Hamiltonian is
\be
H_Q = \sum_a 2(\epsilon_a - \mu) Q_3^{(a)} + 
u_*(N_*/2 + Q_3)^2 + u \vert  Q_+ \vert ^2 \quad ,
\label{sf7}
\ee
where $u_*= U/ N_*$. Two remarks are now in order. First we note that the
long time dynamics is weakly influenced by those k-modes for which
$\vert  \delta_a - \delta_F \vert  > \delta_*$, so that the complexity of
the dynamical behavior now issues from the $Q$-system, as manifestly suggested
by the fact that the restricted $Q$-system has inherited the same structure
the $N$-pseudospins system. This is the main consequence of the adiabatic
approach. Secondly, we recall that the density of states of the
non-interacting system has in two dimensions a logarithmic divergence for
$\epsilon \simeq 0$\cite{HIR} which implies that the levels $\tilde\Lambda_a$
with $\epsilon_a \simeq 0$ are the most populated ones. For situations where
the value of $n$ involves $\epsilon_F\simeq 0$ (near half-filling) such a fact
well matches with the first observation since it turns out that the $Q$-system,
whose dynamics is complex, is also the sub-system involving the most part
of $k$-modes.

 From the above observations, one is led to restricting the number of
interacting levels in order to work out the simplest yet
still significant dynamics. The corresponding model turns
to be a 3-level system, namely the pseudospin model where the $Q$-system is
endowed with three levels. A simple calculation allows one to establish
that the number of constants of motion is not sufficient to make the system
integrable. In this sense 3-level dynamics still is far from being
trivial, yet it is physically meaningful in several circumstances.

At first, for example, one can take into account just the three inner most
levels of the $k$-space, {\sl i.e.} those around $\epsilon_a=0$, which
in the following we shall label by $a = -1, 0, +1$. 
This is natural when investigating the low energy dynamics of ${\cal H}_{sc}$
at half-filling with $u >0$. In this case, in fact, it is reasonable to expect
that increasing the
energy from the ground state value of small amounts (recall that $S_3^{(0)} = 0$
and $S_3^{(\pm)} =\pm N_{\pm 1}/2$ with $N_1 = N_{-1}$) makes interacting just
the levels with the smallest energy, i.e. those with $a = -1, 0, +1$. 
In view of the fact that $\vert  \epsilon_{a+1} - \epsilon_{a} \vert  \ll 4t$,
expressing the almost continuous character of $\epsilon_a$ vs. $a$,
one can replace both the upper level and the lower level of the
3-level model with two sheaves constituted by those levels with
$\epsilon_a \simeq \epsilon_{+1}$ and $\epsilon_a \simeq \epsilon_{-1}$
respectively. This allows one to enlarge the number of modes partecipating
to the dynamics as well as to treat situation where excited states are
more than small perturbations. We recall however that this case is
mainly pedagogic, as at half-filling the ground state of hamiltonian
$H$ has antiferromagnetic order.

Such observations readily extend to those situations where $n \ne 1$.
In these cases, in fact, the minimum energy configuration is not symmetric
with respect to $a = 0$, but with respect to the level $a=F$, where
$S_3^{(a)}$ change from negative to positive so as to minimize the energy.
Then the 3-level construction must be referred to the new central level
thus obtained.

The further reduction to a 2-level scenario immediately makes
the $Q$-system integrable. Again by replacing the two levels with two
$\it effective$ levels one can reasonably expect to still represent the main
features of $Q$-dynamics, in particular when the energy is low enough
to make interacting a limited number of levels situated around the
level with $a = F$.

\section{two-level dynamics}  

In the previous section we noted that the 3-level system is nonintegrable
although it is the oversqueezed version of a multilevel system that
was dramatically more complex. A thorough investigation of its dynamics,
where the occurrence of a chaotic behavior indeed is expected due to its
similitude with the dynamical model of refs. \cite{BOGU}, \cite{BRWG},
requires a separate, extended analysis that will be pursued elsewhere.
Nevertheless we shall start with the 3-level model equations, so as to make
the approximations performed to achieve the 2-level scenario evident.

Let us express the 3-level system equations in the
form given by eqs. (\ref{sf5}) by renaming $S_{\alpha}^{(\nu)}$
($\alpha=+,-,3$) for $\nu=F+1,F,F-1$ by $P_\alpha$, ${\cal Z}_\alpha$, and
$M_\alpha$ respectively, and by setting $\delta_{\pm} = \delta_{F \pm 1}$,
in order to simplify the notation and
to recall the interpretation of the levels as level-sheaves.
The equations then read
\begin{eqnarray}
-i \hbar \dot M_+ &=& \delta_-  M_+ - 2 u M_3 Q_{+}\quad ,\quad
i \hbar \dot M_{3} = u ( M_{+} Q_{-} - M_{-} Q_{+} )\quad ; 
\label{t1} \\
-i \hbar \dot {\cal Z}_+ &=& - 2 u {\cal Z}_3 Q_{+}\quad , \quad 
i \hbar \dot {\cal Z}_{3} = u ( {\cal Z}_{+} Q_{-} - {\cal Z}_{-} Q_{+} ) \quad ;
\label{t2} \\
-i \hbar \dot P_+ &=& \delta_+  P_+ - 2 u P_3 Q_{+}\quad , \quad
i \hbar \dot P_{3} = u ( P_{+} Q_{-} - P_{-} Q_{+} ) \quad .
\label{t3} 
\end{eqnarray}

For $u >0$ the 2-level model is readily obtained by freezing
the ${\cal Z}$-variables at the values ${\cal Z}_3 = 0= {\cal Z}_+$.
From the physical viewpoint, such an assumption is not particularly
restrictive because it allows one to switch on dynamics through $M$-variables
and $P$-variables starting from the ground state configuration. One should
recall, in fact, that $P_+ = M_+  = {\cal Z}_+ = 0$ characterize the ground
state, while the further condition ${\cal Z}_3 = 0$ can be implemented by
suitably selecting $n$ (see the discussion following (\ref{gs})). The fact that
the set ${\delta_a}$ is almost continuous still ensures the possibility of
chosing $n$ almost arbitrarily.

For $u < 0$, instead, the fact that $P_+, M_+, {\cal Z}_+ \ne 0$ in connection
with the minimum energy state, prevents the system from developing a dynamics 
in which ${\cal Z}_+$ and ${\cal Z}_3$ keep the their ground state values.
Thereby the presence of the central level, even if as a nondynamical level
reminescent of 3-level scenary, is prohibited and the central level must be
embodied within one of the two sheaves, unless one is facing the
nonintegrable version of pseudospin dynamics. At this point the 2-level scenario
is restored and one can proceed to integrate the equations of motion.

We construct now the solutions of 2-level dynamics by solving
simultaneously the systems of eqs. (\ref{t1}) and (\ref{t3}).
The main variable of the system is $D_3 = P_3 - M_3$ which will be
shown to obey a nonlinear equation completely decoupled from the other
variables. Indeed the knowledge of $D_3(t)$, together with the constant
of motion $Q_3$ allows one to integrate such system \cite{PERA}, which becomes
linear with time-dependent
coefficients. To work out the equation for $D_3$ one needs to 
exploit all the constants of motion. Explicitly, the energy $H_2
= \delta_+ P_3 + \delta_- M_3 
+ u_{*}(N_{*}/2 + Q_3)^2 + u \vert  P_+ + M_+ \vert ^2$, 
$N_{*} = N_+ + N_-$ counting the active modes of 2-level dynamics,
must be used to eliminate4 the variables $P_{\pm}$ and $M_{\pm}$ from
the equation
\be
{\hbar}^2 {\dot D_3}^2 =4 u^2 \vert  P_{+} M_{-} - P_{-} M_{+} \vert ^2 \quad,
\label{t8}
\ee
obtained via the second equation in (\ref{t1}) and in (\ref{t3}).
This is done by exploiting first the identity
$\vert  P_{+} M_{-} - P_{-} M_{+} \vert ^2 = 
4 \vert  P_{+} \vert ^2 \vert  M_{+} \vert ^2 
-( \vert  P_{+} + M_{+} \vert ^2 -\vert  P_{+} \vert ^2 -\vert  M_{+} \vert ^2
)^2 $, which leads to rewrite the r.h.s. of (\ref{t8}) as
\be
4 u^2 \vert  P_{+} M_{-} - P_{-} M_{+} \vert ^2 =
16 u^2  \vert  P_{+} \vert ^2 \vert  M_{+} \vert ^2 -
4 \Bigl [h_2 -\sigma D_3 - u(\vert  P_{+} \vert ^2 +\vert  M_{+} \vert ^2)
\Bigr]^2\quad,\label{t9} 
\ee
where $ h_2 = H_2 -\gamma Q_3 - u_* (Q_3 + N_*/2)^2$, 
$\gamma = ( \delta_+ + \delta_- )/2$, and $\sigma =(\delta_+ - \delta_- )/2$.
Then, by using the
pseudospin Casimir's $I_M = M_3^2 + \vert  M_{+} \vert ^2$,
$I_P = P_3^2 + \vert  P_{+} \vert ^2$ and the further identities
$P_3^2 + M_3^2 = (Q_3^2 + D_3^2)/2$, $P_3^2 - M_3^2 = Q_3 D_3 $, 
one reduces the $D_3$-equation to the closed form
\be
{\frac{\hbar^2} {2}} {\dot D_3}^2 = 
- 2(h_2 - \sigma D_3)^2 
+ 2u (h_2 - \sigma D_3)(2I - Q_3^2 - D_3^2)
- 2u^2 (I_P - I_M - Q_3 D_3)^2 \quad,
\label{t10} 
\ee
with $I= I_P + I_M$. 

Equation (\ref{t10}), which expresses the integrable
character of the 2-level system, presents several interesting features.
First of all it shows that the $D_3$-dynamics is as complex as that of a
1-d potential problem. Indeed, upon introducing the potential
\bq
{\cal U}(D_3)=  -2 a D_3^3 + 2b D_3^2 + 2c D_3 + 2d  
\nonumber
\eq
where $a = u\sigma$, $b =\sigma^2 + u h_2 + u^2 Q_3^2 $,
$c = u \sigma (2I - Q_3^2)- 2\sigma h_2 - 2u^2 Q_3 (I_P-I_M)$
and $d = (h_2 - uI)^2 + u h_2 Q_3^2 - 4 u^2 I_M I_P$, equation (\ref{t10})
simply becomes $\hbar^2 {\dot D_3}^2 /2 = - {\cal U}(D_3)$.  
Therefore the dynamical behavior
of the 2-level system can be completely specified by identifying
the regions where the cubic ${\cal U}(D_3)$ is negative and finding the value
of the derivative of ${\cal U}(D_3)$ when $D_3$ approaches an inversion point.
Such regions actually are identified by the compact interval in
the potential well of ${\cal U}(D_3)$ whose extremes coincide with two of
the three roots of the cubic equation ${\cal U}(D_3) = 0$. The remaining
semi-infinite interval where ${\cal U}(D_3)$ tends to $-\infty$ must be
excluded in that $D_3$ there would assume infinitely large values,
while its range is finite: $-N_- \le D_3 \le N_+$.

The two other points which play some role in characterizing the dynamics through
${\cal U}(D_3)$ are, of course, the potential stationary points
\bq
R_{\pm} = {\frac {1}{3a}} \Bigl[ b \pm \bigl( b^2 + 3 a c \bigr)^{1/2} \Bigr]
\quad,
\nonumber
\eq
furnished by $d{\cal U} /dD_3 = 0$. In fact, upon denoting the minimum
and the maximum coordinates by $D_m$ and $D_M$ respectively,
it is possible to identify the fixed points as those configurations
of the  2-level
system with ${\cal U}(D_m)=0$ so that the $D_3$-interval reduces to the point
$D_3 = D_m$. It is easily checked that $D_m$ is consistent
with the general result given by (\ref{gs})-(\ref{fixedsc}) for the exact
minimum energy points. On the other hand, when the initial conditions imply
${\cal U}(D_m)< 0$ and ${\cal U}(D_M) >0$, then the system oscillates
between two extreme states. From the physical point of view this
property has the interesting consequence that the filling of
each of the two mesoscopic levels near the Fermi surface ($F\pm 1$) varies
periodically with time, while the sum of the two fillings remains
constant. 

On the other hand, it is worth noticing that such periodic behavior of
the mesoscopic levels filling ceases to exist for appropriate initial
conditions. In fact when ${\cal U}(D_M) = 0$ the
system exhibits a $\it laser$-like effect, namely it tends,
employing an infinitely long time, to an asymptotic stationary state in
which (depending on the sign of $u$) a mesoscopic level is totally
empty while the other is full.
Moreover one should recall that each choice
of the constants of motion $h_2$, $Q_3$, $I_M$, and $I_P$ embodied in $a$, $b$,
$c$, and $d$ selects a different cubic potential. Two solutions are
associated with the same potential when they differ just for the choice of the
initial position $D_3(0)$. 

The nice feature of eq. (\ref{t10}) is that it can be reduced to the equation
for the Weierstrass ${\cal P}$ function \cite{DA}, which reads 
\bq
\Bigl({\frac {d {\cal P}}{d \tau}} \Bigr)^2 = 4{\cal P}^3 - g_2 {\cal P} - g_3
\quad,
\nonumber
\eq
Any solution of (\ref{t10}) can then be given in explicit form.
A straigthforward calculation based on the substitution of $D_3$ with
$D_3 = \pm {\cal P} + b/3a$ ( the plus (minus) corresponds to the case 
$a> 0$ ($a< 0$)) turns (\ref{t10}) into the above equation for
${\cal P}$ where $\tau = {\sqrt a} t/ \hbar$ and the standard
coefficients $g_2$, $g_3$ are identified as
\bq
g_2 = {\frac {4}{3 a^2}}(b^2 + 3ac) \quad,\quad
g_3 = \pm {\frac {4}{27 a^3}}(2b^3 + 9abc + 27 a^2 d)\,\,.
\nonumber
\eq
Then, by exploiting the solution $\cal P$ in terms of Jacobi elliptic
functions \cite{ABST} 
${\cal P}(\tau) = {\frac {1}{3}} \gamma^2 (1 + k^2)
- \gamma^2 k^2 sn^2(\gamma \tau + \alpha ;k)$ where
\bq
g_2 = {\frac{4}{3}} \gamma^4 (1 -k^2 +k^4) \quad, \quad 
g_3 = {\frac{4}{27}} \gamma^6 (1 - 2k^2)(2 - k^2)(1 + k^2) 
\nonumber
\eq
the explicit analitic expression of $D_3(t)$ is easily shown to be\cite{PERA}
\be
D_3(t) =  {\frac {b}{3a}} \pm 
\Bigl[ {\frac {1}{3}} \gamma^2 (1 + k^2)
- \gamma^2 k^2 sn^2(\gamma \tau + \alpha ;k) \Bigr] \,\,.
\label{t11}
\ee

Such solution, as expected, shows that the dynamics is periodic around the
point $D_m$ with period
\bq
T_2 = 2 \hbar {\frac {K(k^2)} {\gamma \vert  a \vert ^{1/2}}}
\nonumber
\eq
since the elliptic sine fulfils the equation $sn(x + 2K) = -sn(x)$, where
$K(k^2)$ is the elliptic integral of first kind \cite{ABST}. As anticipated,
a special case can be selected out when initial conditions allow the condition
${\cal U} (D_M) = 0$ to occur. This condition states that one of the two
${\cal U}$-roots confining the oscillations of $D_3$ inside the potential
well, coincides with the maximum coordinate $D_M$. Reaching $D_M$ thus requires
an infinitely long time provided the motion starts exactly at the other
inversion point, that is the remaining root of ${\cal U}(D_3) = 0$. When this
is the case $k \rightarrow 1$ so that $sn(x,k) \rightarrow th(x)$ and $D_3(t)$
describes indeed a transition for $t \rightarrow \infty$.

In order to esplicitly provide a situation where such transition happens
we concisely examine the dynamics for $u>0$ when $Q_3 =0$ and $I_M = I_P$.
In this case the potential $\cal U$ manifestly exhibits its roots since
it reduces to
\be
{\cal U}(D_3) = - 2 (h_2 - \sigma D_3) 
\bigl( h_2 - 2uI - \sigma D_3 + u D_3^2 \bigr) \quad.
\label{t12} 
\ee 
Hence the circumstance where ${\cal U}(D_3) =0$ for $D_3 = D_M$ is obtained
by imposing the two roots of the quadratic factor in (\ref{t12}) to merge
i.e. to tend to $D_M$. As a result one finds first the constraint
$4u h_2 = \sigma^2 + 8 u^2 I$ on the energy, then that the limiting point of
the transition is $D_3(\infty) = \displaystyle{\frac {\sigma}{2u}} =D_M$ when
$D_3(0)= h_2/\sigma$, provided the condition
$4{\sqrt{I_P}}< \vert  \sigma \vert  / u   < 2N_+$ holds ensuring that
$-N_+ < D_M < D_3(0)$.

Returning to (\ref{t11}), one can now easily evaluate the phase
$\varphi_{sc}(t)$ given by expression (\ref{phasesc}) in the simplest case
in which $S_+^{(k)} = S_+^{(a)} / N_a$ for any $k$. Since this involves that
$S_3^{(k)} = \pm S_3^{(a)} / N_a$, then pseudospins populating a level at most
differ one from the other by the sign of $S_3^{(k)}$.
We shall choose the same sign for the $k$'s of a given level so that
the microscopic dynamics is just a copy of the mesoscopic dynamics.
Here, we shall not consider the possibility
of more structurated configurations of pseudospins, since they do not introduce
any substantial novelty concerning the phase behavior. 

Finally, by using the above assumption, $\varphi_{sc}(t)$ can be written in the
form
\be
{\dot \varphi}_{sc} = -{\cal H}_{sc} + \hbar \sum_a (N_a/2 + S_3^{(a)})
{\dot \lambda}_a
\quad.
\label{ph1}
\ee
The latter expression is particularly useful for the dynamics of weakly
excited states since the variables $S_3^{(a)}$'s are expected to undergo small
variations in time, with respect to their ground state values. The levels
whose pseudospins are fast should therefore contribute to $\varphi_{sc}$
principally through the phases $\dot \lambda_a$'s. On the other hand, after
solving eqs. (\ref{sf3}) for the pseudospins labelled by
$a= F+2$ and $a= F-2$ (namely pseudospins of the first two levels with fast
variables) so as to have a 4-level system mimicking the real system,   
a simple calculation shows the time average of ${\dot \lambda}_{F\pm 2}$ to be
almost zero in the $Q$-time scale. Indeed the slow variables $M_3$ and $P_3$
and the phases of $M_+$ and $P_+$ provide the main contribution to
${\dot \varphi}_{sc}$ even when the fast variables are included.
The expression of such contribution can be readily obtained by rewriting
(\ref{ph1}) in terms of $M_3$, $P_3$, and of the constants of motion.
One thus finds
\be
\dot\varphi_{sc}\simeq -{\cal H}_{sc} +  u \Biggl \{ n^2 {{N^2}{4}}
-4(M_+ P_- + M_- P_+){\frac {N_-P_3 - N_+ M_3}{(N_+ -2P_3)(N_- -2M_3)}}
\Biggr\}
\quad.
\ee
The slow variation in time of its variables makes it a good candidate for
experimental detection.

\section{Equations of motions in the antiferromagnetic phase}

When the possibility of a AF phase is considered, the natural order
parameter which has to be non zero is \cite{HIR}
\be
m\doteq{1\over N}<\sum_{{\bf j}\in \Lambda}
{\rm e}^{i {\bf G}\dot {\bf j}} (n_{{\bf j},\uparrow}-n_{{\bf
j},\downarrow})> = {1\over N} <\sum_{{\bf k}\in \tilde\Lambda}
a_{{\bf k},\uparrow}^\dagger a_{{\bf k}-{\bf G},\uparrow}-
a_{{\bf k},\downarrow}^\dagger a_{{\bf k}-{\bf G},\downarrow}>
\quad ,
\ee
where ${\bf G}$ is a vector with all its components equal to $\pi$.

Also here, we look at the reduced Hartree-Fock Hamiltonian in the AF
phase, $H_l^{(af)}$, in order to derive the dynamical algebra in which we shall
subsequently construct our GCS's. It reads
\bq
H_l^{(af)}=-t \sij\sum_\sigma \cisd\cjs + {U\over 2} m \sum_{\bf j}
{\rm e}^{i {\bf G} {\bf j}} (\nup-\ndw)-U{N\over 4}(1-m^2)\nonumber 
\eq
and it can be rewritten in reciprocal space as $H_l^{(af)}= \sum_{{\bf k}
\in\tilde\Lambda_-,\sigma}H_{{\bf k},\sigma}^{(af)}$, where now 
\bq
H_{{\bf k},\sigma}^{(af)} =\epsilon_{\bf k} \sum_\sigma
(n_{{\bf k},\sigma}-n_{{\bf k}-{\bf G},\sigma})+U {m\over 2} (
a_{{\bf k},\uparrow}^\dagger a_{{\bf k}-{\bf G},\uparrow}-
a_{{\bf k},\downarrow}^\dagger a_{{\bf k}-{\bf
G},\downarrow}+h.c.)+{U\over 4} (m^2-n^2) \quad , \nonumber
\eq
and $\tilde \Lambda_-$ is that half of $\tilde\Lambda$ in which 
$\epsilon_{\bf k}$ is negative, {\sl e.g.} for $d=2$ 
$\tilde\Lambda_-\equiv\{{\bf k}\in \tilde\Lambda \vert  \epsilon_{\bf k}< 0,
{\rm or}\, \epsilon_{\bf k}=0\,{\rm and}\, 0<k_1\leq \pi\}$.

$H_l^{(af)}$ can be recognized as an element of the dynamical algebra
${\cal A}_{af}=\dsp{\bigoplus_{{\bf k}\in\tilde\Lambda_+,\sigma}
{\cal A}_{{\bf k},\sigma}^{(af)}}$, with
\be
{\cal A}_{{\bf k},\sigma}^{(af)} = \left \{ K_{{\bf k},\sigma}^{(+)}
=a_{{\bf k}-{\bf G},\sigma}^\dagger a_{{\bf k},\sigma},
K_{{\bf k},\sigma}^{(-)}=[K_{{\bf k},\sigma}^{(+)}]^\dagger,
K_{{\bf k},\sigma}^{(Z)}={1\over 2} (n_{{\bf k}-{\bf G},\sigma}-
n_{{\bf k},\sigma})\right\} \sim su(2)_{{\bf k,\sigma}} \quad , 
\ee
and $K_{{\bf k},\sigma}^{(\pm)}=K_{{\bf k},\sigma}^{(X)}\pm i
K_{{\bf k},\sigma}^{(Y)}$.
In full analogy with the case treated in section III, we use as 
trial approximate time-dependent wave-function for studying the
full Hamiltonian $H$ in antiferromagnetic phase the time-dependent
generalization of GCS's which can be built in ${\cal A}_{af}$,
{\sl i.e.}
\be
\vert \psi (t)>_{af}={\rm e}^{{i\over\hbar}\varphi_{af}(t)}\vert \xi(t)> = 
{\rm e}^{{i\over\hbar}\varphi_{af}(t)} \Pi_{{\bf k}\in\tilde\Lambda_-,\sigma}
(1+\bar\xi_{{\bf
k},\sigma} \xi_{{\bf k},\sigma})^{-{1\over 2}} \exp{\left(
\xi_{{\bf k},\sigma} K_{{\bf k},\sigma}^{(+)}\right)}\vert 0>_{af}
\quad , \label{psiaf}
\ee
where $\vert 0>_{af}\equiv\Pi_{{\bf k}\in\tilde\Lambda_-,\sigma}\vert \sigma> 
\Pi_{{\bf k}\in\tilde\Lambda_+,\sigma}\vert 0>$ ($\tilde\Lambda_+=
\tilde\Lambda-\tilde\Lambda_-$), and the parameters $\xi_{{\bf k},
\sigma}$ have to be thought of as time-dependent.

The semiclassical Hamiltonian ${\cal H}_{af}$, identified as the expectation
value of $H$ over $\vert \psi (t)>_{af}$, , can be fruitfully
rewritten in terms of the semiclassical variables
$\zeta_{{\bf k},\sigma}\doteq _{af}<\psi(t)\vert K_{{\bf k}\sigma}^{(\zeta)}
\vert \psi(t)>_{af}$, with $\zeta=X,Y,Z$, which still satisfy a $su(2)$
algebra (\ref{su2}). Also here it is useful to introduce a one-dimensional
index $a$ instead of ${\bf k}$, and to define the mesoscopic variables
$\zeta_{a,\sigma}\doteq \sum_{{\bf k}\in\tilde\Lambda_a}
\zeta_{{\bf k}, \sigma}$. One obtains
\be
{\cal H}_{af} = - 2 \sum_{a\in\tilde\Lambda_-,\sigma} \epsilon_a
Z_{a,\sigma}+4 u X_\uparrow X_\downarrow - u{N^2\over 4}\quad ,
\label{hamspinaf}
\ee
with $\zeta_\sigma =\sum_{a\in\tilde\Lambda_+}
\zeta_{a,\sigma}$.

The semiclassical equations of motion for the $\zeta_{a,\sigma}$'s,
can be easily derived from (\ref{hamspinaf}), (\ref{caneq}), and read
\bq
\hbar \dot X_{a,\sigma} &=& 2 \epsilon_a Y_{a,\sigma}\quad ,\nonumber \\
\hbar \dot Y_{a,\sigma} &=& - 2 \epsilon_a X_{a,\sigma}- 4 u
X_{-\sigma} Z_{a,\sigma}\quad ,\nonumber \\
\hbar \dot Z_{a,\sigma} &=& 4 u X_{-\sigma} Y_{a,\sigma}
\quad . \label{eqmotaf}
\eq
It is interesting to notice that the above equations do reduce to
equations formally identical with those studied for the SC-paramagnetic
phases for the special choice $Y_\sigma=0$, $X_\sigma=\pm X_{-\sigma}=
\dsp{1\over 2}S_\pm$, so that at least in this case the dynamics can be
derived from that obtained there. Moreover, let us notice that the choices
$X_{a,\uparrow}= \pm X_{a,\downarrow}\quad , \quad Y_{a,\uparrow}= \pm
Y_{a,\downarrow}$ reduce to a half the number of equations (\ref{eqmotaf}).
One can easily verify that such choices
minimize the value of ${\cal H}_{af}$ in the positive ($-$) and
negative ($+$) $u$ regime respectively.

From eqns. (\ref{eqmotaf}) we also obtain the time dependent phase
charachteristic of the TDVP approach,
\bq
\dot\varphi_{af} &=& -{\cal H}_{af} + \sum_{{\bf k} \in \tilde\Lambda_-}
\frac{\dot Y_{{\bf k},\sigma} X_{\sigma,({\bf k})} -
\dot X_{\sigma,({\bf k})} Y_{{\bf k},\sigma}}{1-2
Z_{{\bf k},\sigma}} \nonumber \\ &=& \phi -4 u
\sum_\sigma\left[X_{-\sigma} \sum_{{\bf k}\in\tilde\Lambda_-}
X_\sigma^{({\bf k})}\frac{1+2 Z_{{\bf k},\sigma}}{ 1 -
2 Z_{{\bf k},\sigma}}\right ] \quad  \label{tdpaf} \quad , 
\eq 
where $\phi=u\dsp{N^2\over 4}-2\sum_{{\bf k} \in \tilde\Lambda_-}
\epsilon_{\bf k}$ is a constant. As in the superconducting case,
also in (\ref{tdpaf}) the time-dependent part of
$\dot\varphi_{af}$ is vanishing for vanishing $u$ as well as
for $X_\sigma=0$, which is related to the vanishing
of the antiferromagnetic order parameter $X$ ($X\doteq
X_\uparrow-X_\downarrow$).

As in the case treated in the previous sections, also here we 
first look for the fixed points of eqs. (\ref{eqmotaf}). A first 
solution is of course the vanishing one, {\sl i.e.}  $X_{a,\sigma} 
=Y_{a,\sigma}=0$, and $Z_{a,\sigma}$ fixed by initial conditions. In
particular, the configuration of $Z_{a,\sigma}$ minimizing the energy 
has energy $E_{af}^{(0)}= -u \dsp{N^2\over 4} +2 \sum_{a\in\tilde\Lambda_-}
\epsilon_a N_a$, 
which is easily verified to coincide with that proper of the paramagnetic
phase (\ref{gs}) in the positive $u$ regime at half-filling.

The remaining set of fixed points can be parametrized by the Casimir's
${\cal I}_{a,\sigma}=X_{a,\sigma}^2+Z_{a,\sigma}^2$, which again are
conserved quantities. It is characterized by $Y_{a,\sigma}=0$, and
\be
X_{a,\sigma}=2 s_{a,\sigma} u\sqrt{{\cal I}_{a,\sigma}\over { \epsilon_a^2 +4
u^2 X_{-\sigma}^2}} X_{-\sigma}\quad , \quad
Z_{a,\sigma}=\epsilon_a s_{a,\sigma} \sqrt{{\cal I}_{a,\sigma}\over
{ \epsilon_a^2+4 u^2 X_{-\sigma}^2}} \quad , \label{fpaf}
\ee
where $s_{a,\sigma}=\pm 1$. The $X_\sigma$'s have to satisfy the
constraint equations $X_\sigma = 2 u \sum_a s_{a,\sigma}
\sqrt{{\cal I}_{a,\sigma}\over { \epsilon_a^2 +4u^2 X_{-\sigma}^2}}
X_{-\sigma}$. The corresponding energy $E_{af}$ reads
\be
E_{af}= -u{N^2\over 4}+\sum_{a,\sigma} \epsilon_a^2 s_{a,\sigma}
\sqrt{{\cal I}_{a,\sigma}\over { \epsilon_a^2 +4u^2 X_{-\sigma}^2}}+
4 u X_\uparrow X_\downarrow \quad . \label{efpaf}
\ee
In particular, the fixed points which minimize (\ref{efpaf}) are
associated with the choices $s_{a,\sigma}=-1$, 
${\cal I}_{a,\sigma}={N_a^2\over 4}$, and $X_\uparrow=-sgn(u) X_\downarrow$.
In this case the constraint equations, apart from the solution $X_\sigma
= 0$, reduce to one,
{\sl i.e.} 
\be
1= \vert u\vert  \sum_a {N_a\over\sqrt{\epsilon_a^2 +4u^2 X_\uparrow^2}} \quad ,
\label{selfaf}
\ee
and the minimum energy is straightforwardly obtained from
(\ref{efpaf})-(\ref{selfaf}) as
\be
E_{af}^{(m)}= - u {N^2\over 4}-2 \sum_{a\in\tilde\Lambda_-} N_a
\sqrt{\epsilon_a^2 +4u^2 X_\uparrow^2} + 4 \vert u\vert  X_\uparrow^2 \quad .
\label{emaf}
\ee

As expected, such energy corresponds to a non-vanishing
antiferromagnetic order parameter $X=2 X_\uparrow$ only for $sgn(u)=+$ 
({\sl i.e.} repulsive Coulomb interaction), whereas it gives $X=0$
for $sgn(u)=-$. In the first case, the
energy $E_{af}^{(m)}$ coincides in fact with the one obtained
within Hartee-Fock approximation, with $X$ replaced by $m$
satisfying the same self-consistency
equation (\ref{selfaf}). On the contrary,
in the attractive Coulomb interaction regime, even though $X=0$
the energy $E_{af}^{(m)}$ is lower than the one obtained within
Hartree-Fock approximation, which would be precisely given by
$E_{af}^{(0)}$,
namely the energy corresponding to the trivial vanishing fixed-point. 
This is not surprising, in that while within the Hartree-Fock scheme the only
parameter to be fixed self-consistently is $m$, here to all effect we have two
related parameters, $X_\uparrow$ and $X_\downarrow$, which
can separately be non-zero even when their difference ({\sl i.e.}
$X$) is vanishing. Recalling that $X_\sigma=_{af}<\psi\vert \sum_{\bf j} 
(-)^{\bf j} n_{{\bf j},\sigma}\vert \psi>_{af}$, this latter case
($X_\uparrow=X_\downarrow\neq 0$) can be recognized as a
CDW phase.

Notice that in the absolute minimum energy point for $u<0$ both the conditions
which reduce the equations of motion (\ref{eqmotaf}) to those of
the superconducting case (\ref{eqmotsc}) were fulfilled (see the
discussion following eq. (\ref{eqmotaf})). A
direct comparison with the result obtained for the negative $u$ regime
by means of the superconducting states (\ref{emsc}) shows
that in fact at half-filling $E_{sc}^{(-)}\equiv E_{af}^{(m)}$.
Hence we derived within TDVP at $u<0$ two degenerate
wave-functions for the ground state, the superconducting and the
charge-density-wave one. Indeed it is easily verified that
the two wave-functions are orthogonal, and that the expectation
value of the order operator of one phase, when taken over the
wave-functions of the other phase, is identically vanishing.

Now let us analyze the equation of motions (\ref{eqmotaf}) away from
the fixed points in some simple case. A first integrable case is
obtained when the variable $X_\sigma$ is kept constant. However this
assumption is consistent only if $\sum_a \epsilon_a  Y_{a,\sigma} =0$,
and such condition in turn is satisfied only if $Y_{a,\sigma}$ is
independent of time for $\epsilon_a\neq 0$. Then the solution
for each $a\neq F$ reduces to (\ref{fpaf}), whereas for $a=F$ 
it turns out to be given by
\be
Y_{F,\sigma}= A_{\sigma} \cos (\alpha_\sigma t) + B_{\sigma}
\sin (\alpha_\sigma t)\quad ,\quad
Z_{F,\sigma}= A_{\sigma} \cos (\alpha_\sigma t) - B_{\sigma}
\sin (\alpha_\sigma t) \quad , \label{sol0af}
\ee
and $X_{F,\sigma}= X_\sigma-\sum_{a\neq F} X_{a,\sigma}$,
with $\alpha_\sigma= 4 \dsp{u\over \hbar} X_{-\sigma}$.
Solution (\ref{sol0af}) survives in correspondence to stationary points of
the Hamiltonian (when $X_\sigma$ are chosen according to the self-consistency
equations), as the system energy is not changed by the value of
$Z_{F,\sigma}$ and $Y_{F,\sigma}$. Such solution describes a periodic
behavior of the mesoscopic Fermi level, holding even
for the interacting ground state. In fact, due to the
Casimir costraint the constants $A_{\sigma}$ and $B_{\sigma}$ turn out
to be related by the equation $A_\sigma^2+B_\sigma^2={\cal I}_{F}-
X_{F,\sigma}^2$. The latter condition implies that $A_\sigma=B_\sigma=0$
for $X_{F,\sigma}=\pm \sqrt{{\cal I}_{F}}$, which is the case only 
for the absolute minimum point of the non-interacting case
(see (\ref{fpaf})). On the contrary, for any
$\vert X_{F,\sigma}\vert <\sqrt{{\cal I}_{F}}$ from (\ref{sol0af}) we
obtain this oscillatory-like behavior of the solution at the
Fermi surface. Such a behavior affects neither the order
parameter, nor the energy, but it turns out to affect the phase
$\varphi_{af}(t)$ characteristic of the TDVP approach, by adding to the
term linear in time a structured periodic time-dependent contribution given by
\be
\sum_{{\bf k}\in\tilde\Lambda_F} \tan^{-1} \left [ {B_{{\bf k},\sigma}
+ A_{{\bf k},\sigma}
\tan \left (\dsp{\alpha_\sigma\over 2} t \right )}\over{\alpha_\sigma
\sqrt{1-4 X_{{\bf k}^*,\sigma}^2}}\right ] \quad . \label{tdp0}   
\ee
Here we used the same dynamics for the local and the mesoscopic
pseudospin variables. In summary, we obtained also in the AF and CDW
phases a non-trivial phase dynamics for the ground state.

Apart from this simple case, more generally the system described by
(\ref{eqmotaf}) has been investigated in the case were the mesoscopic
levels which have fast dynamics are one or two\cite{BOGU,BRWG}. Already
within such framework it appears to have very interesting chaotic
properties. 

\section{Conclusions}

In the present paper we developed a consistent scheme for
dealing with the dynamics of an itinerant interacting many-electron system
described by the Hubbard Hamiltonian. Such scheme is based on TDVP
procedure, and has been applied for describing the dynamics of the model
by means of macroscopic wave-functions built in terms of GCS 
of the dynamical algebras which generate the Hartree-Fock
solution in SC, AF, CDW, and paramagnetic phases.
Already for these simple cases a certain number of remarkable features
related to the dynamical description rather than to the statistical-mechanical
one was underlined.

First of all, a geometric phase --a macroscopic quantity which in principle
is observable-- occurs for appropriate values of the physical parameters in
the ground state as well as for some low-energy excited states . Such feature
can not be identified by solving the eigenvalue equation for the
Hamiltonian (or related techniques, like the Bethe Ansatz approach),
as it is a consequence of the phase of the eigenfunction, which
in the eigenvalue equation is free. Even more noticeably, away from
half-filling in the repulsive regime, it was shown that
such macroscopic behavior of the Berry phase is originated from
a vortex-like dynamics of the phases of the microscopic variables. 
Both these features could be due to the approximations implied by our
scheme, hence a first interesting point which is left open to future
work is to study exactly the dynamics of the stationary points of $H$,
by solving the Schr\"odinger equation near them. This could be done by
using the Glauber GCS's, which map exactly the quantum Hamiltonian into
its semiclassical form, and studying the fixed points of the resulting
equations of motion \cite{MOPE}.

Other interesting dynamical properties of the system were stressed
in the low-energy regime for some integrable cases. At half-filling,
the ground state has been shown to exhibit an oscillatory-like behavior
at the Fermi surface. Away from half-filling, for $u>0$ 
an analogous oscillating behavior for the mesoscopic density variable
takes place near the Fermi surface.
Such a feature is responsible for
a non-trivial time dependence of the collective Berry phase. Again,
this point should be further analyzed in different approximations.
An alternative viewpoint could be furnished even
by employing the same TDVP scheme starting from GCS's more realistic than
the Hartree-Fock ones. For instance, in the $U\rightarrow\infty$ limit,
a reliable basis is given by the Gutzwiller states \cite{GUT}. 

One more solution obtained exactly within the present scheme and exhibiting
interesting features is the single-mode solution, characterized by the
non-vanishing of the superconducting parameter, and by a unique
time-dependent phase reflecting collective order.
The possible relevance of this solution within the framework
of superconductivity is related to the fact that it survives at
$u>0$, and at any energy but the ground state, with a volume
in the space of solutions increasing with energy.

All the above solutions, which are exact within the present approximation
scheme, are interesting also in that they could represent good starting points
for studying more exhaustively the dynamics described by (\ref{eqmotsc})
in their neighbourhood, by means of standard perturbative methods of classical
dynamics. As a general conclusive observation let us notice that their
validity beyond the present TDVP scheme could be tested  by 
solving exactly the Schr\"odinger equation for $H_{Hub}$ on small clusters of
sites. Work is in progress along these lines.

\acknowledgments

The authors thank Mario Rasetti for helpful comments and critical reading
of the manuscript.


\begin{references}
\bibitem{BMT} R.A. Broglia, A. Molinari, and T. Regge, Ann. of Phys.
{\bf 109}, 349 (1977)
\bibitem{PER}A.M. Perelomov, {\it Generalized Coherent States and their
Applications}, (Springer Verlag, Berlin, 1986)
\bibitem{ZFG}W.M. Zhang, D.H. Feng, and R. Gilmore, Rev. Mod. Phys. {\bf
62}, 867 (1990)
\bibitem{FUTS}T. Fukui, and Y. Tsue, Progr. Theor. Phys. {\bf
87}, 627 (1992)
\bibitem{BRLO}H.B. Braun, and D. Loss, Phys. Rev. B {\bf 53}, 3237 (1996)
\bibitem{TITI}D.R. Tilley, and J. Tilley, {\it Superfluidity and
superconductivity}, (Adam Hilger LTD, Bristol and Boston, 1986) 
\bibitem{HOHA}J.R. Hook, and H.E. Hall, {\it Solid State Physics},
(John Wiley and Sons, Chichester, 1991)
\bibitem{HUB}J. Hubbard, Proc. R. Soc. London A {\bf 276}, 
238 (1963) and {\bf 277}, 237 (1964); M.C. Gutzwiller, {it Phys. Rev. Lett.} 
{\bf 10}, 159 (1963).
\bibitem{SHWI} A. Shapere, and F. Wilczek, {\it Geometric phases in
physics}, (World Scientific, Singapore, 1992)
\bibitem{AHAN} Y. Aharonov, and J. Anadnan, Phys. Rev. Lett {\bf
58}, 2281 (1987) 
\bibitem{BER} M.V. Berry, Proc. R. Soc. London A{\bf 392}, 45
(1984) 
\bibitem{MOPE} A. Montorsi, and V. Penna, in preparation
\bibitem{MON}{\it The Hubbard model: a reprint volume}, A. Montorsi ed.
(World Scientific, Singapore, 1993).
\bibitem{LIE} E. H. Lieb, ``The Hubbard model -- Some Rigorous Results and 
Open Problems,'' in {\it Proceedings of the XIth
International Congress of Mathematical Physics}, D. Iagolnitzer
ed., (International Press, Diderot 1995).
\bibitem{LIWU} E.H. Lieb, and F.Y. Wu, Phys. Rev. Lett. {\bf 20}, 1445 
(1968).
\bibitem{HIR} J.E. Hirsh, Phys. Rev. B {\bf 31}, 4403 (1985)
\bibitem{LIMO} R. Livi, A. Montorsi, and M. Rasetti, Mod. Phys.
Lett. B{\bf 5}, (1992)
\bibitem{ARPL}D.K. Arrowsmith, and C.M. Place, {\it An Introduction to
Dynamical Systems}, (Cambridge University Press, Cambridge, 1990)
\bibitem{SAKU}H. Sakaguchi, and Y. Kuramoto, Progr. Theor. Phys. 
{\bf 76}, 576 (1986)
\bibitem{KU}Y. Kuramoto, Progr. Theor. Phys. {\bf 79}, 223 (1984)
\bibitem{ME} A. Messiah, {\it Quantum Mechanics}, North-Holland, Amsterdam, 1962

\bibitem{BOGU} L.L. Bonilla and F. Guinea, Phys. Rev. {\bf 45}, 7718 (1992)

\bibitem{BRWG} L. Bonci, R. Roncaglia, B.J. West, and P. Grigolini,
Phys. Rev. {\bf 45}, 8490 (1992)

\bibitem{PERA} V. Penna, and M. Rasetti, Phys. Rev. B {\bf 50}, 11783 (1994)

\bibitem{DA} H.T. Davis, {\it Introduction to Nonlinear Differential and
Integral Equations}, (Dover, New York, 1970)

\bibitem{ABST} {\it Handbook of mathematical functions} M. Abramowitz,
and I. Stegun eds. (Dover, New York, 1972)

\bibitem{GUT}M.C. Gutzwiller, Phys. Rev. {\bf 137}, A1726 (1965)

 
\end{references}
\end{document}